\documentclass[prb,twocolumn]{revtex4-1}

\usepackage{amsmath,amsfonts,amssymb}
\usepackage{graphicx,color}
\usepackage{hyperref}
\usepackage{epstopdf}
\usepackage{float}
\usepackage{multirow}
\begin{document}

\title{Multi-mode superconducting circuits for realizing strongly coupled multi-qubit processor units}

\author{Tanay Roy$^{1}$, Madhavi Chand$^{1}$, Anirban Bhattacharjee$^{1}$, Sumeru Hazra$^{1}$, Suman Kundu$^{1}$}
\author{Kedar Damle$^{2}$}
\author{R. Vijay$^{1}$}

\affiliation{$^{1}$Department of Condensed Matter Physics and Materials Science, Tata Institute of Fundamental Research, Homi Bhabha Road, Mumbai 400005, India}
\affiliation{$^{2}$Department of Theoretical Physics, Tata Institute of Fundamental Research, Homi Bhabha Road, Mumbai 400005, India}

\date{\today}

\begin{abstract}
Inter-qubit coupling and qubit connectivity in a processor are crucial for achieving high fidelity multi-qubit gates and efficient implementation of quantum algorithms. Typical superconducting processors employ relatively weak transverse inter-qubit coupling which are activated via frequency tuning or microwave drives. Here, we propose a class of multi-mode superconducting circuits which realize multiple transmon qubits with all-to-all longitudinal coupling. These ``artificial molecules" directly implement a multi-dimensional Hilbert space that can be easily manipulated due to the always-on longitudinal coupling. We describe the basic technique to analyze such circuits, compute the relevant properties and discuss how to optimize them to create efficient small-scale quantum processors with universal programmability. 

\end{abstract}

\maketitle

\section{Introduction}

Superconducting circuits have revolutionized experiments in quantum mechanics due to the flexibility offered in constructing designer Hamiltonians by appropriately combining linear inductors, capacitors and Josephson junctions. Apart from being a leading candidate for building quantum computers \cite{sup-qubit-review-science}, superconducting circuits have led to tremendous progress in the field of microwave quantum optics \cite{qoptics_rev}, ultra-low noise amplification \cite{paramp_review,amplifying_vacuum_review} and hybrid quantum devices \cite{hybrid_sc} as well. For applications in quantum computing, continuous enhancement in the circuit design over the past two decades has led to the improvement of coherence time from nanoseconds \cite{nakamura_cpb} to milliseconds \cite{fluxonium_ms_manucharyan,fluxonium_ms_schuster}. The transmon qubit \cite{shunted-flux-qubit, transmon_theory}, the most popular design used for multi-qubit experiments, evolved from the modification of the Cooper pair box (CPB) qubit \cite{nakamura_cpb} by carefully tailoring the Josephson and charging energy to provide significant resilience to charge noise while retaining sufficient anharmonicity for fast gate operations. An alternate approach used inductive shunting of the CPB and led to the fluxonium qubit \cite{fluxonium}, which suppressed the effects of charge noise while retaining the strong non-linearity of the CPB qubit. More recently, significant improvements in flux qubit coherence \cite{Flux-qubit-coherence, fluxqubit_new} have been achieved by intelligent modification of circuit design and parameters.

A major challenge in building larger scale quantum processors is high fidelity multi-qubit gates and inter-qubit connectivity in a processor. A majority of the multi-qubit architectures have used individual transmon qubits which are transversely coupled to each other using nearest neighbor capacitances or bus cavities \cite{Martinis_9xmon_nature, IBM-errordet-4qubit, riste-bit-flip, multi-qubit-3d-yale,Qsim_xmon, wallraff-dig-qsim, RIP-gate}. This typically leads to weak inter-qubit coupling with restricted inter-qubit connectivity. We recently introduced the trimon \cite{trimon}, a multi-mode superconducting circuit implementing a strongly coupled three-qubit system with all-to-all connectivity. Such circuits can be thought of as ``artificial molecules" \cite{fluxonium_molecule} where the individual qubits are so strongly coupled that they lose their original identity and hybridized modes emerge. While previous work on such multi-mode circuits has mostly focused on creating better effective qubits or enabling tuning of qubit properties \cite{Houck-TCQ2,Buisson-Vshaped,houck-TCQ-photonnoise,fluxonium_molecule}, we propose the use of such multi-qubit units as building blocks for quantum processors with the possibility of higher fidelity gates and better inter-qubit connectivity \cite{trimon}.

In this article, we first outline a general scheme for analyzing multi-mode circuits comprising of strongly coupled anharmonic oscillators and show how it leads to all-to-all longitudinal coupling\cite{nature-cpb-schnirman, cpb-array-nori, Auffeves_Buisson-theory} between the emergent transmon like qubits. We then focus on the properties of a three-qubit multi-mode device (trimon) and describe how to deal with non-idealities, optimize parameters for multi-qubit operations and perform joint readout of the multi-qubit state. We conclude by discussing possible design variations and some practical considerations for robust operation of such circuits.

\section{Formalism}
\label{sec:formalism}

\begin{figure}[t]
	\centering
	\includegraphics[width=0.9\columnwidth]{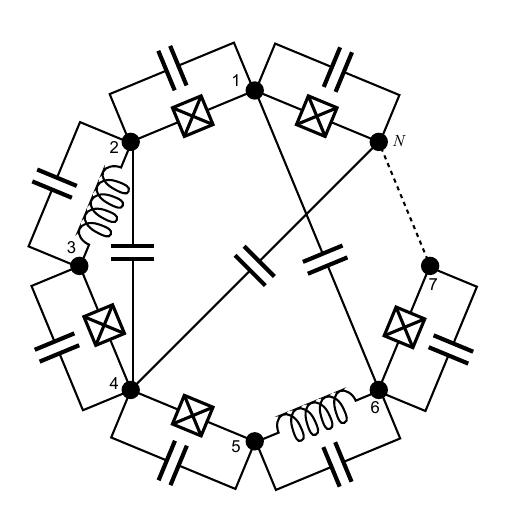}
	\caption{Schematic of an $N$-node system connected by an arbitrary set of Josephson junctions, linear inductors and capacitors between pairs of nodes. Each node always has a small capacitance $C_{ii}$ to the ground (not shown in the figure). We do not consider any other type of circuit element between the nodes and ground.}
	\label{fig:N-node}
\end{figure}

We consider an $N$-node system with arbitrary connections of non-dissipative lumped elements in between pairs of nodes as shown in Fig.~\ref{fig:N-node}. The elements between nodes $i$ and $j$ can consist of a linear inductor (with inductance $L_{ij}$), a linear capacitor (with capacitance $C_{ij}$) and a Josephson junction (with Josephson energy ${E_J}_{ij}$). We determine the normal modes and inter-mode coupling of the circuit by adapting the formalism described in Refs.\cite{bbq-yale,quantize-blais, bbq-divincenzo} and describe it as a multi-qubit system in the circuit-QED architecture. Note that the treatment described here is valid only when the emergent normal modes can be considered as weakly nonlinear oscillators with anharmonicities similar or lower than those of transmon qubits \cite{transmon_theory}. This constraints the value of Josephson energies that can be used and also requires that each node is shunted to at least one other node with a large enough capacitor. Further, we mostly consider operation at zero flux through any closed superconducting loops and freeze the fluxon degree of freedom. The effect of magnetic flux in specific circuits is discussed later in Section \ref{sec:ring}.

 We begin our analysis by defining the inductive energy matrix and the capacitance matrix as,
\begin{equation}
\label{eq:EJmat}
	E_L=
	\begin{pmatrix}
	\sum_i{E_L}_{1i} & -{E_L}_{12} & -{E_L}_{13} & \cdots &  -{E_L}_{1N} \\
	-{E_L}_{21} & \sum_i{E_L}_{2i} & -{E_L}_{23} & \cdots &  -{E_J}_{2N} \\
	\vdots &  \vdots & \vdots & \ddots &  \vdots \\
	-{E_L}_{N1} & -{E_L}_{N2} & -{E_L}_{N3} & \cdots &  \sum_i{E_L}_{Ni} \\
	\end{pmatrix},
\end{equation}
\begin{equation}
C=
\begin{pmatrix}
\sum_i C_{1i} & -C_{12} & -C_{13} & \cdots &  -C_{1N} \\
-C_{21} & \sum_i C_{2i} & -C_{23} & \cdots &  -C_{2N} \\
\vdots &  \vdots & \vdots & \ddots &  \vdots \\
-C_{N1} & -C_{N2} & -C_{N3} & \cdots &  \sum_i C_{Ni} \\
\end{pmatrix},
\end{equation}
where ${E_L}_{ij} = \varphi_0^2/L_{ij} + {E_J}_{ij}$ is the inductive energy arising due to the linear inductance and Josephson junction connected between nodes $i$ and $j$ with $\varphi_0=\hbar/2e$ being the reduced flux quantum. Assuming that there is no external flux threading any closed superconducting loops in the circuit (see section \ref{sec:ring}),  the Lagrangian of the linearized system can be expressed as 
\begin{equation}
\label{eq:Lagrangian}
\mathcal{L}=\dfrac{1}{2}\sum_{i,j=1}^N \left[ \dfrac{d\Phi_i}{dt} C_{ij} \dfrac{d\Phi_j}{dt} - {E_L}_{ij} \left( \dfrac{\Phi_i-\Phi_j}{\varphi_0} \right)^2 \right],
\end{equation}
where $\Phi_i$ are the node fluxes of each node related to the potentials as $V_i=\dot{\Phi_i}$. 

Finding the normal modes of this system requires simultaneous diagonalization of the capacitance and inductive energy matrices. This is possible because the capacitance matrix is always positive definite for a physical system (ground capacitances $C_{ii}$ are always non-zero even if small). 
Simultaneous diagonalization is done by first determining the eigenvalues ${\lambda_C}_\mu$ and orthonormal eigenvectors ${v_C}_\mu$ of the capacitance matrix $C$. Next one rewrites the node fluxes in terms of new variables $\Phi_j=\sum_{\mu=0}^{N-1}\mathfrak{X}_\mu v_{C_{\mu j}} / \sqrt{\lambda_{C_\mu}}$ where $v_{C_{\mu j}}$ represents the $j$-th element of the $\mu$-th eigenvector $v_{C_{\mu}}$. In terms of these variables the Lagrangian becomes
\begin{equation}
\label{eq:lagrangian_unitmass}
\mathcal{L}=\dfrac{1}{2} \left[ \sum_{\mu=0}^{N-1}\left( \dfrac{d\mathfrak{X}_\mu}{dt} \right)^2 - \sum_{\mu,\nu=0}^{N-1} {\tilde {E_L}}_{\mu \nu} \mathfrak{X}_\mu \mathfrak{X}_\nu \right],
\end{equation}
where the matrix $\tilde{E}_L=\tilde{C}^T E_L \tilde{C}$. Here $\tilde{C}$ is a matrix whose columns are the eigenvectors ${v_C}_\mu$ divided by $\sqrt{{\lambda_C}_\mu}$. Eq. \ref{eq:lagrangian_unitmass} is the Lagrangian of a system of $N$ coupled oscillators all having unit mass. This can now be diagonalized by finding eigenvalues ${\lambda_L}_\mu$ and orthonormal eigenvectors $\Xi_\mu$ of the matrix $\tilde{E}_{L_{\mu \nu}}$. Mode frequencies $\omega_\mu$ are then related to  ${\lambda_L}_\mu$ as $\omega_\mu=\sqrt{{\lambda_L}_\mu}$ and the matrix $\Xi$ whose columns are composed of eigenvectors $\Xi_\mu$, relate the node-fluxes $\Phi_{i=1,\cdots,N}$ with mode-flux variables $\tilde{\Phi}_\mu$ as
\begin{equation}
\label{eq:mode-structure}
\Phi_i=\sum_{\mu=0}^{N-1} (\tilde{C}\Xi)_{i\mu} \tilde{\Phi}_\mu , \ \ {i=1,2,\cdots,N}.
\end{equation}
In terms of the mode-flux variables ($\tilde{\Phi}_\mu$) which represent the normal modes of the system, the Lagrangian reads
\begin{equation}
\label{lagrangian_modeflux}
\tilde{\mathcal{L}}=\dfrac{1}{2} \sum_{\mu=0}^{N-1} \left[ \left( \dfrac{d\tilde{\Phi}_\mu}{dt} \right)^2 - \omega_\mu^2 \tilde{\Phi}_\mu^2 \right].
\end{equation} 
We use the convention $\tilde{\Phi}_{\mu=0}$ to describe the zero frequency mode which can be considered as a charging mode. Since we are working in the weakly anharmonic oscillator limit, there are no charging effects and an excitation of this mode will not affect the other non-zero frequency modes. As a result, $\tilde{\Phi}_{\mu=0}$ does not couple to the other modes and can thus be ignored in the analysis that follows.

After solving the linear model, one can find the non-linear properties of the original system by substituting the mode-structures from Eq.~\eqref{eq:mode-structure}  into the full potential $U=f(\Phi_1,\Phi_2,\cdots,\Phi_N)$, which in general can take any functional form provided the non-linearity can be treated as a perturbation. Restricting ourselves to linear inductors and Josephson junctions, the potential energy becomes
\begin{equation}
U=\sum_{i>j=1}^N \left[ \dfrac{{E_L}_{ij}}{2} \left(\dfrac{\Phi_i-\Phi_j}{\varphi_0}\right)^2 - {E_J}_{ij} \cos\left(\dfrac{\Phi_i-\Phi_j}{\varphi_0}\right) \right].
\end{equation}
After transforming to normal mode coordinates ($\tilde{\Phi}_\mu$), quantization of the system\cite{Kerr-coeff} is achieved by substituting,
\begin{equation}
\tilde\Phi_\mu\rightarrow \sqrt{\dfrac{\hbar}{2 \omega_\mu}}\left( a_\mu + a_\mu^\dagger \right),
\end{equation}
where $a_\mu (a_\mu^\dagger)$ is the bosonic annihilation (creation) operator for the $\mu$-th mode. Application of rotating wave approximation (i.e., keeping the energy conserving terms only) on the system leads to harmonic terms of the type $(a_\mu^\dagger a_\mu)$ along with leading-order nonlinear terms of the type $(a_\mu^\dagger a_\mu)^2$ and $(a_\mu^\dagger a_\mu)(a_\nu^\dagger a_\nu)$ whose coefficients (negative for circuits with Josephson junctions) determine the strength of self-Kerr $(J_\mu)$ and cross-Kerr $(J_{\mu \nu})$ type nonlinearity respectively. Usually, it is sufficient to consider the lowest few orders of $\tilde\Phi_\mu$ (say up to ${\tilde\Phi}_\mu^6$) in the expansion of $U$ as the contributions from higher order terms become insignificant. At this stage, one should verify that the obtained self-Kerr terms are small compared with the mode frequencies ($\lesssim 5\%$) to ensure the validity of the weakly anharmonic approximation used throughout this calculation. The self-Kerr term causes a gradual shift in the transition frequencies as one climbs up the ladder of energy eigenstates for a particular mode while the cross-Kerr term establishes pairwise coupling between two modes. We call this class of devices as the ``multimon" which behave as multiple transmon qubits with pairwise longitudinal coupling.

\section{Ring Multimon Devices}
\label{sec:ring}

\begin{figure}[t]
	\centering
	\includegraphics[width=\columnwidth]{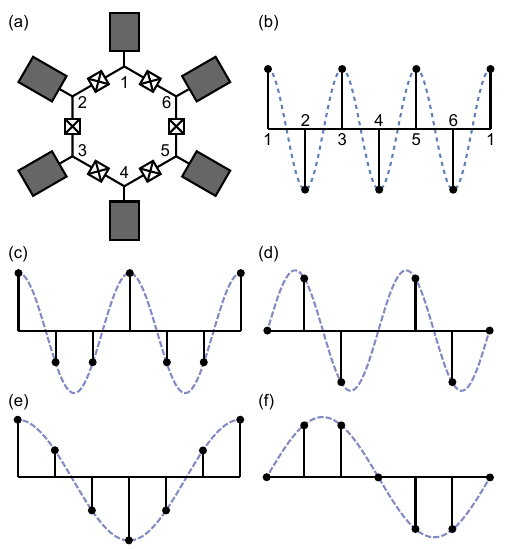}
	\caption{(a) Ring multimon with 6 identical Josephson junctions and capacitor pads in 3D geometry which provide the inter-node capacitance for the multimon. (b)-(f) Mode structures for the 5 orthogonal modes which can be visualized as standing waves on a discrete string with periodic boundary condition - (b) octupolar mode, (c), (d) quadrupolar modes and (e), (f) dipolar modes. The solid black circles represent relative amplitudes of the node fluxes at each node. }
	\label{fig:pentamon}
\end{figure}

In this section, we analyze a specific circuit geometry where Josephson junctions are placed only between nearest-neighbor nodes while capacitors connect every pair of nodes. The trimon device (see section \ref{subsec:trimon}) we introduced in Ref.~\cite{trimon} belongs to this category of ``ring multimon" devices. In this case the inductive energy matrix takes the bi-diagonal form (except for the corner elements) since ${E_J}_{ij}\neq 0$ only if $|i-j|=1$. A realization in 3D geometry of a six-node ring device  is shown in Fig.~\ref{fig:pentamon}(a), where all Josephson junctions and capacitor pads are identical and placed symmetrically about the center. In the case of an $N$-node ring with $N$-fold rotational symmetry (henceforth called the symmetric ring multimon), the $N-1$ non-trivial (finite frequency) eigenmodes can be visualized as standing waves on a discrete string with periodic boundary conditions (Fig.~\ref{fig:pentamon}). The mode-shapes of a symmetric ring multimon thus can be expressed by the following vectors,
\begin{equation}
\label{eq:ideal-modes}
\tilde{\varphi}_\mu=
\begin{cases}
   \{ \sin \left( \dfrac{2\pi j \lceil \mu/2 \rceil}{N} \right) \}, \ \mu\in even, \ j\in(0,N-1) \\
   \{ \cos \left( \dfrac{2\pi j \lceil \mu/2 \rceil}{N} \right) \}, \ \mu\in odd, \ j\in(0,N-1)
\end{cases}
\end{equation}
where $\lceil \ \rceil$ is the ceiling function and $\tilde{\Phi}_\mu \propto \tilde{\varphi}_\mu$. Fig.~\ref{fig:pentamon}(b-f) show the geometric mode structures for the case of a symmetric six-node device depicted in Fig.~\ref{fig:pentamon}(a). If $N$ is an even number, the $N$-fold rotationally symmetric ring multimon provides pairwise degenerate modes except for the highest frequency-mode (Fig.~\ref{fig:pentamon}(b)). In order to use an $N$-node ring as a system of $N-1$ spectroscopically distinct qubits, one needs to break the rotational symmetry by introducing asymmetry in the junctions or capacitors or both. In the presence of any asymmetry in the system, the emergent modes will become linear superpositions of the vectors given in Eq.~\eqref{eq:ideal-modes} and in general require numerical solution.

Using an appropriate choice of device parameters, the modes can be made to act as transmon qubits with all-to-all longitudinal coupling described by the Hamiltonian
\begin{equation}
\label{eq:Hsys}
 \begin{split}
 \dfrac{1}{\hbar}H_{\rm{sys}} &= \sum_{\mu=1}^{N-1} \left[ (\omega_\mu - \beta_\mu) a_\mu^\dagger a_\mu - J_\mu (a_\mu^\dagger a_\mu)^2 \right] \\
 & - \sum_{\mu\neq\nu} 2J_{\mu \nu} (a_\mu^\dagger a_\mu)(a_\nu^\dagger a_\nu)\\ 
 & + \sum_{\mu\neq\nu\neq\zeta} J_{\mu\nu\zeta}(a_\mu^\dagger a_\mu)(a_\nu^\dagger a_\nu)(a_\zeta^\dagger a_\zeta),
 \end{split}
\end{equation} 
where,
\begin{equation}
\beta_\mu = J_\mu + \sum_{\nu\neq \mu}J_{\mu \nu}.
\end{equation}
Usually, the three-body coupling terms $J_{\mu\nu\zeta}$ are two orders of magnitude smaller than the two-body coupling terms $J_{\mu\nu}$ and can be ignored for all practical purposes.

In the presence of a finite external flux $\Phi_{\rm{ext}}$ through the loop, there will be additional phase drop across each junction due to the static current flowing through the ring. As a result the potential energy in Eq.~\eqref{eq:Lagrangian} has to be modified to
\begin{equation}
U=\sum_{i=1}^N {E_J}_{i,i+1} \cos \left( \dfrac{\Phi_i-\Phi_{i+1}+\Phi_{i,i+1}^{\rm{DC}}}{\varphi_0} \right),
\end{equation}
where the values of $\Phi_{i,i+1}^{\text{DC}}$ have to be found by numerically solving the following equations
\begin{subequations}
	\begin{align}
	{E_J}_{12}\sin\left( \dfrac{\Phi_{12}^{\rm{DC}}}{\varphi_0} \right) = {E_J}_{23}\sin\left( \dfrac{\Phi_{23}^{\rm{DC}}}{\varphi_0} \right)\\
	\vdots\\
	{E_J}_{N-1,N}\sin\left( \dfrac{\Phi_{N-1,N}^{\rm{DC}}}{\varphi_0} \right) = {E_J}_{N1}\sin\left( \dfrac{\Phi_{N1}^{\rm{DC}}}{\varphi_0} \right)\\
	\sum_{i=1}^N \Phi_{i,i+1}^{\rm{DC}}=\Phi_{\rm{ext}}
	\end{align}
\end{subequations}
External flux modifies the frequencies of the orthogonal modes and introduces three-body transverse coupling terms $\tilde\Phi_\mu \tilde\Phi_\nu \tilde\Phi_\zeta$ which remain ineffective unless a resonant condition (three-wave mixing \cite{JRM}) is met. However, qubit anharmonicities $(-2J_\mu)$ and longitudinal coupling strengths $(J_{\mu \nu})$ remain mostly unaffected. This implies that one can tune the mode frequencies down as long as the highest energy eigenstate remains stable (see section \ref{subsec:levels}). 

Placing a ring multimon inside a cavity resonator leads to coupling of every mode to the cavity with varying strength since each mode, in general, has a dipolar component along the cavity's electric field. However, only the two dipolar modes of a symmetric ring multimon (equivalent to Fig.~\ref{fig:pentamon}(e-f)) can couple to the cavity depending upon their relative orientation with respect to the cavity's electric field. The Hamiltonian of this multi-mode system can be expressed as an extended version of the Jaynes-Cummings model
\begin{equation}
\label{eq:HJC}
 \begin{split}
 \dfrac{1}{\hbar}H_{\rm{JC}} &= \sum_{\mu=1}^{N-1} \left[ (\omega_\mu - \beta_\mu) a_\mu^\dagger a_\mu - J_\mu (a_\mu^\dagger a_\mu)^2 \right] \\
 & - \sum_{\mu\neq\nu} 2J_{\mu \nu} (a_\mu^\dagger a_\mu)(a_\nu^\dagger a_\nu) + \omega_r(a_r^\dagger a_r+1/2)\\ 
 & + \sum_{\mu=1}^{N-1} g_\mu(a_\mu^\dagger a_r + a_\mu a_r^\dagger),
 \end{split}
\end{equation} 
where $a_r \ (a_r^\dagger)$ is the bosonic annihilation (creation) operator for the cavity mode with resonant frequency $\omega_r$. Relative amplitudes of the dipolar coupling strengths $g_\mu$ for each mode $\mu$ can be computed numerically by taking a projection of the mode-vector in the direction of the cavity's field. Assuming all the modes are far detuned from the cavity, i.e., $|\omega_\mu-\omega_r| \gg g_\mu$, we can apply the dispersive approximation to Eq.~\eqref{eq:HJC} to obtain
\begin{equation}
\label{eq:Hdisp}
\begin{split}
\dfrac{1}{\hbar}H_{\rm{disp}} &= \sum_{\mu=1}^{N-1} \left[ (\omega_\mu - \beta_\mu) \hat{n}_\mu - J_\mu (\hat{n}_\mu)^2 \right] \\
& - \sum_{\mu\neq\nu} 2J_{\mu \nu} \hat{n}_\mu \hat{n}_\nu + \omega_r(\hat{n}_r+1/2)\\ 
& + \sum_{\mu=1}^{N-1} g_\mu^2 \bigg( \dfrac{\hat{n}_\mu(\hat{n}_r+1)}{\Delta_\mu(\hat{n}_A,\hat{n}_B,\cdots,\hat{n}_{N-1})+2J_\mu} \\
& \ \ \ \ \ \ - \dfrac{(\hat{n}_\mu +1)\hat{n}_r}{\Delta_\mu(\hat{n}_A,\hat{n}_B,\cdots,\hat{n}_{N-1})} \bigg),
\end{split}
\end{equation} 
where $\hat{n}_\mu=a_\mu^\dagger a_\mu$ is the photon number operator for mode $\mu$ and
\begin{equation}
\Delta_\mu(\hat{n}_A,\hat{n}_B,\cdots,\hat{n}_M)=\Delta_{\mu 0}-2J_\mu \hat{n}_\mu - \sum_{\nu\neq\mu}2J_{\mu\nu} \hat{n}_\nu ,
\end{equation}  with $\Delta_{\mu 0} =\omega_\mu-2J_\mu-\sum_{\nu\neq \mu}J_{\mu\nu}-\omega_r$. The dependence of the effective detuning $\Delta_\mu$ (between mode $\mu$ and the cavity) on the occupation of all other modes is a consequence of all-to-all longitudinal coupling.

Since $H_{\rm{disp}}$ is diagonal in the photon number basis, the energy of the full system can be obtained by simply substituting the photon number operators with the corresponding occupation numbers in the Hamiltonian. Let us represent the energy of the total system by $E_{\rm{tot}}(n_A,n_B,\cdots,n_{N-1},\hat{n}_r)$ and restrict ourselves to single photon excitations in the individual ring-modes since we are interested in qubit operations only. The effective resonator frequency $\tilde{\omega}_r$ will in general depend upon the occupation of all qubits. However, the dispersive shift for single qubit excitation can be found by measuring the change in cavity's frequency when a particular qubit jumps from ground state ($n_\mu=0$) to excited state ($n_\mu=1$), while keeping all other qubits in their ground states. The total shift $2\chi_\mu$ for qubit $\mu$ can be extracted by looking at the coefficient of $n_r$ in $E_{\rm{tot}}(n_A=0,n_B=0,\cdots,n_\mu=1,\cdots,n_{N-1}=0,n_r) - E_{\rm{tot}}(n_A=0,n_B=0,\cdots,n_\mu=0,\cdots,n_{N-1}=0,n_r)$ leading to
\begin{equation}
\begin{split}
\chi_\mu &=g_\mu^2 \left( \dfrac{1}{\Delta_{\mu 0}}-\dfrac{1}{\Delta_{\mu 0}-2J_\mu} \right) \\
&+ \sum_{\nu\neq \mu}\dfrac{g_\nu^2}{2} \left( \dfrac{1}{\Delta_{\nu 0}}-\dfrac{1}{\Delta_{\nu 0}-2J_{\mu \nu}} \right).
\end{split}
\label{eq:disp_shift}
\end{equation}
The dispersive shift in Eq.~\eqref{eq:disp_shift} has two components - the first part is the standard dispersive shift \cite{transmon_theory} coming from direct qubit-cavity coupling, whereas the second term is the indirect pull on the cavity via the inter-qubit longitudinal coupling.

By appropriately choosing the symmetry of the ring multimon, some of the direct coupling strengths ($g_\mu$) can be made zero, making those modes protected from Purcell decay \cite{Blais_TCQ}. Even these protected modes possess finite dispersive shift coming from the second term in Eq.~\eqref{eq:disp_shift} or in other words all $N-1$ qubits are measurable via a dispersive shift \cite{trimon}. 

\section{Trimon}
\label{subsec:trimon}
\begin{figure}[t]
	\centering
	\includegraphics[width=\columnwidth]{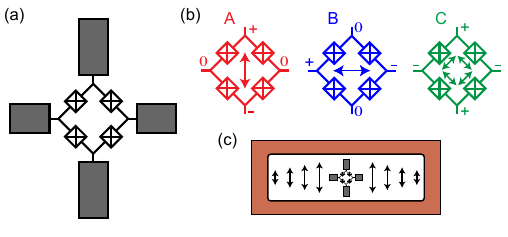}
	\caption{(a) Schematic of a trimon qubit (in 3D geometry) with four identical Josephson junctions and diagonally symmetric capacitor pads which provide the inter-node capacitances for the trimon. (b) Three modes of the trimon - two dipolar modes shown by blue and red arrows and a quadrupolar mode shown by green arrows. (c) Placement of the trimon chip inside a copper cavity with qubit A's dipole aligned with the cavity's electric field for the TE$_{101}$ mode.}
	\label{fig:trimon-cav}
\end{figure}

We now focus on the four-junction multimon device, called the trimon \cite{trimon}, which has three orthogonal modes \cite{JRM} (Fig.~\ref{fig:trimon-cav}) acting as three longitudinally coupled transmon qubits. Recently, such a device has been proposed as a coupling element between two qubits for applications in quantum annealing \cite{Transmon-annealer-Zoller}. In the case of identical junctions and capacitor pads the trimon has two degenerate dipolar qubits (A and B) and one quadrupolar qubit (C) with higher frequency. The degeneracy between qubits A and B can be lifted by symmetrical modification of the diagonal capacitor pads (Fig.~\ref{fig:trimon-cav}(a)) making the three qubits addressable by their frequencies. 
The mode structures of the trimon can be expressed by the following vectors
\begin{equation}
\begin{split}
\tilde{\varphi}_A &= \{ 1, 0, -1, 0 \},\\
\tilde{\varphi}_B &= \{ 0, 1, 0, -1 \},\\
\tilde{\varphi}_C &= \{ 1, -1, 1, -1 \},
\end{split}
\label{eq:trimon_vector}
\end{equation} 
where the elements represent relative node-fluxes at each node of the ring. If the trimon is placed inside an electromagnetic cavity in a way that the fundamental (TE$_{101}$) mode of the cavity is aligned with the dipole of qubit A, then qubit A becomes strongly coupled. Qubits B and C being orthogonal to A, ideally remain uncoupled from the cavity and hence protected from Purcell decay \cite{Blais_TCQ}. The Hamiltonian of the system in the dispersive limit can be expressed as,
\begin{equation}
 \begin{split}
 \dfrac{1}{\hbar}H_{\rm{sys}} &= \sum_{\mu=A,B,C} \left[ (\omega_\mu - \beta_\mu) \hat{n}_\mu - J_\mu (\hat{n}_\mu)^2 \right] \\
 & - \sum_{\mu\neq\nu} 2J_{\mu \nu} \hat{n}_\mu \hat{n}_\nu + \omega_r(\hat{n}_r+1/2)\\ 
& + g_A^2 \bigg( \dfrac{\hat{n}_A(\hat{n}_r+1)}{\Delta_A(\hat{n}_A,\hat{n}_B,\hat{n}_C)+2J_A} \\
& \ \ \ \ \ \ \ \ \ - \dfrac{(\hat{n}_A +1)\hat{n}_r}{\Delta_A(\hat{n}_A,\hat{n}_B,\hat{n}_C)} \bigg),
 \end{split}
\end{equation}
with effective (single excitation) dispersive shifts,
\begin{subequations}
\begin{align}
	\chi_A &= g_A^2 \left( \dfrac{1}{\Delta_{A0}} - \dfrac{1}{\Delta_{A0} - 2J_A} \right),\\
	\chi_B &= \dfrac{g_A^2}{2} \left( \dfrac{1}{\Delta_{A0}} - \dfrac{1}{\Delta_{A0}-2J_{AB}} \right),\\
	\chi_C &= \dfrac{g_A^2}{2} \left( \dfrac{1}{\Delta_{A0}} - \dfrac{1}{\Delta_{A0}-2J_{CA}} \right),
\end{align}
\end{subequations} 
where $\Delta_{A0}=\omega_A-2J_A-2J_{AB}-2J_{CA}-\omega_{r}$ is the effective detuning between qubit A and cavity. Note that qubit A modifies the cavity's frequency through the direct coupling \cite{transmon_theory}, whereas qubits B and C pull the cavity's frequency indirectly by modifying qubit A's frequency through the inter-qubit coupling \cite{tcq-houck,Blais_TCQ}. We observe that for typical parameters, the dispersive shifts for all three qubits are similar. This property of measurability along with Purcell protection makes B and C almost ideal qubits. However, in practice, the uncoupled nature of qubits B and C makes them very difficult to excite. A small,  controlled asymmetry in the junctions can help alleviate this problem without sacrificing Purcell protection completely. At the same time, any real device also comes with some spread in the Josephson energies due to fabrication uncertainties. We now discuss the effect of asymmetry in the device on its properties.

\subsection{Junction and capacitor asymmetry}
\label{subsec:asymmetry}
\begin{figure}[t]
	\centering
	\includegraphics[width=\columnwidth]{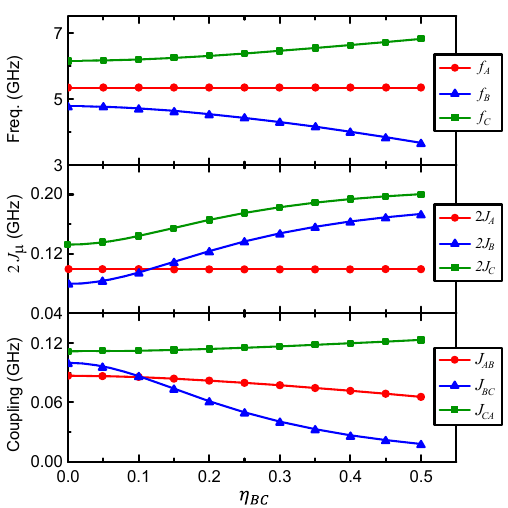}
	\caption{Variation of transition frequencies (top panel), anharmonicities (middle panel) and inter-qubit coupling (bottom panel) as a function of asymmetry parameter  $\eta_{BC}$ for ${E_J}_m=9$ GHz, $\ C_{12}=C_{23}=C_{34}=C_{41}=36$ fF, $C_{13}=12$ fF, and $\ C_{24}=24$ fF. The transition frequency and anharmonicity of qubit A remain unchanged since mixing occurs between original B and C modes. The coupling strength between qubits B and C changes dramatically while it remains mostly unaffected between the other pairs. }
	\label{fig:asym}
\end{figure}

\begin{figure}[t]
	\centering
	\includegraphics[width=\columnwidth]{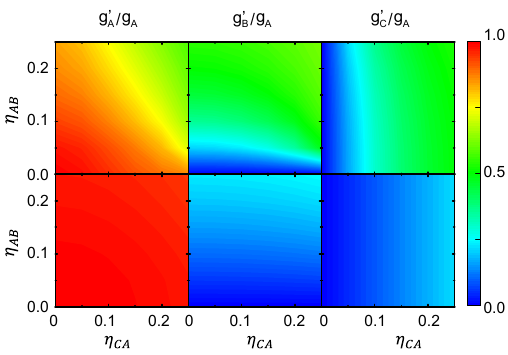}
	\caption{Variation of direct coupling $(g_\mu')$ of the three qubits as a function of different asymmetries for ${E_J}_m=9$ GHz, $\ C_{12}=C_{23}=C_{34}=C_{41}=36$ fF, $C_{13}=12$ fF, and $\ C_{24}=24$ fF. The top panel displays results when only junction asymmetries $\eta_{AB}$ and $\eta_{CA}$ are varied. As the junction asymmetries are increased, extent of mixing increases resulting in increased direct coupling for qubits B and C and reduced coupling for qubit A. The bottom panel shows the compensation of junction asymmetries by choosing capacitance asymmetries equal to the junction asymmetries, i.e.,  $\eta'_{AB}=\eta_{AB}$ and $\eta'_{CA}=\eta_{CA}$. This choice almost suppresses the mixing between B and C modes and brings the mode structures back to their original form as shown in Fig.~\ref{fig:trimon-cav}(b). Note that further optimization to compensate junction asymmetries is possible using numerical techniques.}
	\label{fig:gasym}
\end{figure}

The asymmetry in the junctions can be parametrized using three numbers, which determine which two modes mix. Let $\eta_{\mu \nu}$ be the relative coefficient of asymmetry which mixes modes $\mu$ and $\nu$. We can then write the Josephson energies of the four junctions in terms of the asymmetry parameters as:
\begin{subequations}
	\begin{align}
	{E_J}_{12} &= {E_J}_m(1+ \eta_{AB} +\eta_{BC} +\eta_{CA}),\\
	{E_J}_{23} &= {E_J}_m(1- \eta_{AB} +\eta_{BC} -\eta_{CA}),\\
	{E_J}_{34} &= {E_J}_m(1+ \eta_{AB} -\eta_{BC} -\eta_{CA}),\\
	{E_J}_{41} &= {E_J}_m(1- \eta_{AB} -\eta_{BC} +\eta_{CA}),
	\end{align}
\end{subequations}
where ${E_J}_m$ is the mean of the Josephson energies of the four junctions. Asymmetry $\eta_{BC}$ maintains the node and anti-node locations for the A mode leaving it unaffected, while mixing  modes B and C. This results in increased anharmonicity of modes B and C and reduced inter-qubit coupling. In the presence of any asymmetry, the mode-structure and other parameters can only be determined numerically. The variation of frequencies, anharmonicities and couplings as a function of $\eta_{BC}$ are shown in Fig.~\ref{fig:asym}.
Other asymmetries behave in a similar fashion keeping a particular mode largely unaffected. The introduction of two or more types of asymmetries causes all the three modes to become linear combinations of original dipolar and quadrupolar modes represented by Eq.~\eqref{eq:trimon_vector}. As a result, each mode develops a dipolar component which directly couples to the cavity making them easier to excite at the cost of reduced Purcell protection. The direct coupling strength of each mode $g'_\mu$ can be estimated using the following expression 
\begin{equation}
\label{eq:gprime}
g_\mu'=g_A {\tilde\Phi_\mu'^T \tilde\Phi_A}/{|\tilde\Phi_A|^2},
\end{equation}
where $\tilde\Phi_\mu' = \sum_{i=1}^{N} V_{\mu i}' {\Phi}_i$ are the new normal modes obtained by following the normal-mode analysis described in section \ref{sec:formalism}, and $|\tilde\Phi_A|$ denotes the norm of vector $\tilde{\Phi}_A$. Eq.~\eqref{eq:gprime} essentially computes the projection of the modified $\mu$ mode on the original A mode vector which was directly coupled to the cavity with strength $g_A$.  The variation in direct coupling of the three qubits as a function of junction asymmetries $\eta_{AB}$ and $\eta_{CA}$ (while keeping $\eta_{BC}=0$) are shown in the top panel of Fig.~\ref{fig:gasym}. As a result, in the presence of finite asymmetry, the dispersive shift of each qubit gets contribution from both direct and indirect coupling. The general expressions for computing dispersive shifts for all eight energy eigenstates are given in Appendix \ref{app:dispersive}.
It is clear from Figs.~\ref{fig:asym}, \ref{fig:gasym} and Eq.~\eqref{eq:chi_full} that the junction asymmetries provide some flexibility in targeting device parameters according to the experimental needs.

Asymmetry in the capacitances can also be modeled in a similar fashion:
\begin{subequations}
	\begin{align}
	{C}_{12} &= {C}_m(1+ \eta_{AB}' +\eta_{BC}' +\eta_{CA}'),\\
	{C}_{23} &= {C}_m(1-\eta_{AB}' +\eta_{BC}' -\eta_{CA}'),\\
	{C}_{34} &= {C}_m(1+\eta_{AB}' -\eta_{BC}' -\eta_{CA}'),\\
	{C}_{41} &= {C}_m(1-\eta_{AB}' -\eta_{BC}' +\eta_{CA}'),
	\end{align}
\end{subequations} 
where $C_m$ is the mean of all nearest-neighbor capacitances and capacitance asymmetry coefficients $\eta_{\mu\nu}'$ play similar roles as $\eta_{\mu\nu}$. It is interesting to note that junction asymmetries can be partially compensated by capacitor asymmetries and provides another knob for tuning device parameters. The cancellation effect can be seen in the bottom panel of Fig.~\ref{fig:gasym} which depicts the relative coupling strength of each mode when capacitance asymmetries $\eta'_{AB}$ and $\eta'_{CA}$ are made identical to junction asymmetries $\eta_{AB}$ and $\eta_{CA}$. In this case, qubits B and C hardly develop any direct coupling to the cavity and the coupling of qubit A remains almost unmodified.

\subsection{Level spacing and flux tuning}
\label{subsec:levels}
\begin{figure}[t]
	\centering
	\includegraphics[width=1\columnwidth]{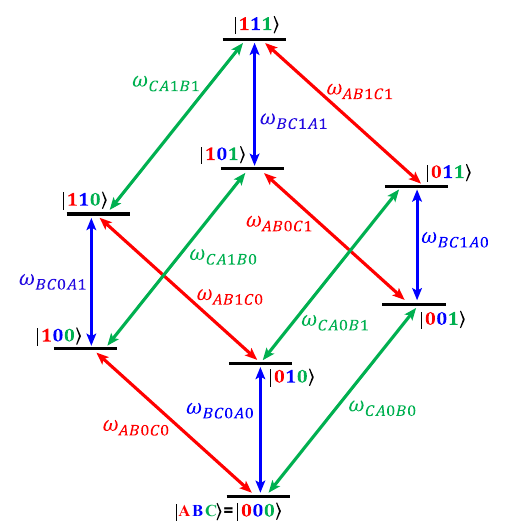}
	\caption{Energy level diagram of the computational subspace for a trimon showing eight energy eigenstates with twelve transitions. The red, blue and green arrows represent transitions belonging to qubits A,B and C respectively, e.g., $\omega_{AB0C1}$ is the transition of qubit A when qubit B is in $|0\rangle$ and qubit C is in $|1\rangle$. For a practical device the parameters should be chosen in such a way that these transitions are well separated from each other in order to achieve fast gate operations.}
	\label{fig:level}
\end{figure}
In order to utilize the trimon as a coupled three-qubit system, the levels containing up to single excitation in every mode are used as the computational subspace. Fig.~\ref{fig:level} displays the energy level diagram for a trimon device with eight energy eigenstates and twelve transitions (each qubit having four conditional transition frequencies). It is important to choose the device parameters in such a way that these transitions are well separated from each other for fast gate operations. Further, we need to ensure that the $|1\rangle \rightarrow |2\rangle$ transition for each mode does not come in close proximity to the transitions within the computational subspace to minimize leakage errors. By appropriately adjusting the device parameters and/or introducing more Josephson junctions between diagonal nodes one can move these unwanted transitions away. Another important criterion is to ensure that the absolute energy of the $|111\rangle$ level be considerably smaller than the $4{E_J}_{\rm{min}}=4\times\min{[{E_J}_{12},{E_J}_{23},{E_J}_{34},{E_J}_{41}]}$. Otherwise, the presence of saddle points in the potential landscape at a height of $4{E_J}_{\rm{min}}$ can significantly perturb the state $|111\rangle$ and can make it unstable \cite{flurin_thesis}. While designing devices with more than three coupled qubits is straightforward (Fig.~\ref{fig:pentamon}), we observe that the anharmonicities and coupling strengths tend to reduce in magnitude with increasing number of junctions in the ring. Such a device will suffer from frequency crowding, slower gates and will pose more stringent conditions on device parameters for optimal level spacing. 

\begin{figure}[t]
	\centering
	\includegraphics[width=\columnwidth]{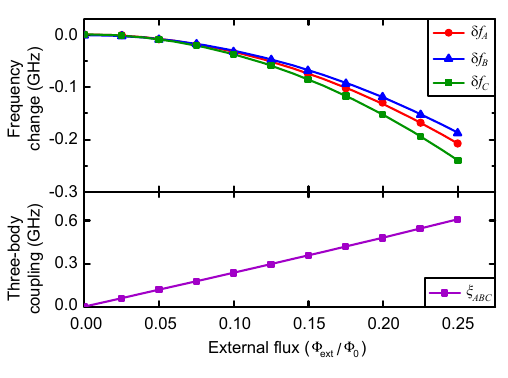}
	\caption{Flux-tuning of the trimon. The top panel shows change in transition frequencies $\delta f_{\mu=A,B,C}$ from corresponding zero-flux values (5.338, 4.778 and 6.156 GHz  for qubits A,B and C respectively) and the bottom panel displays three-body transverse coupling for a trimon with ${E_J}_m=9$ GHz, $\ C_{12}=C_{23}=C_{34}=C_{41}=36$ fF, $C_{13}=12$ fF, and $\ C_{24}=24$ fF. Here $\Phi_0=h/2e$ is the magnetic flux-quantum. The qubit frequencies decrease gradually with magnetic-flux due to reduction in the effective Josephson energies. The anharmonicities and inter-qubit longitudinal coupling strengths remain almost unaffected (change by $<0.2\%$) and are not shown. The three-body transverse coupling ($\xi_{ABC}$) increases rapidly with flux due to increasing asymmetry in the potential, but has no adverse effect unless the three-wave mixing condition is satisfied.}
	\label{fig:flux}
\end{figure}

Introduction of magnetic flux in the trimon loop causes a gradual reduction in qubit frequencies $(f_\mu)$ while leaving anharmonicites $(-2J_\mu)$ and couplings $(J_{\mu\nu})$ almost unchanged.  However, the three-body transverse coupling $(\xi_{ABC})$ increases rapidly as shown in Fig.~\ref{fig:flux}. The reason behind this effect is that the magnetic flux only modifies the Josephson energies of individual junctions and not the charging energies which control the anharmonicity and inter-qubit coupling. Since $\xi_{ABC}$ is ineffective unless a resonance condition is satisfied \cite{JRM}, the trimon behaves as a flux-tunable three-qubit system. However, one should be careful not to apply flux greater than a quarter flux-quantum as the system might become susceptible to phase slips \cite{phase-slip_chain,phase-slip_rhombi,flurin_thesis} and hence become unstable. Further, the approximation of weak nonlinearity used in the current treatment becomes less accurate with increasing flux.

\subsection{Multi-qubit gate operations}
\label{subsec:multiqubitgates}
\begin{figure*}[t]
	\centering
	\includegraphics[width=\textwidth]{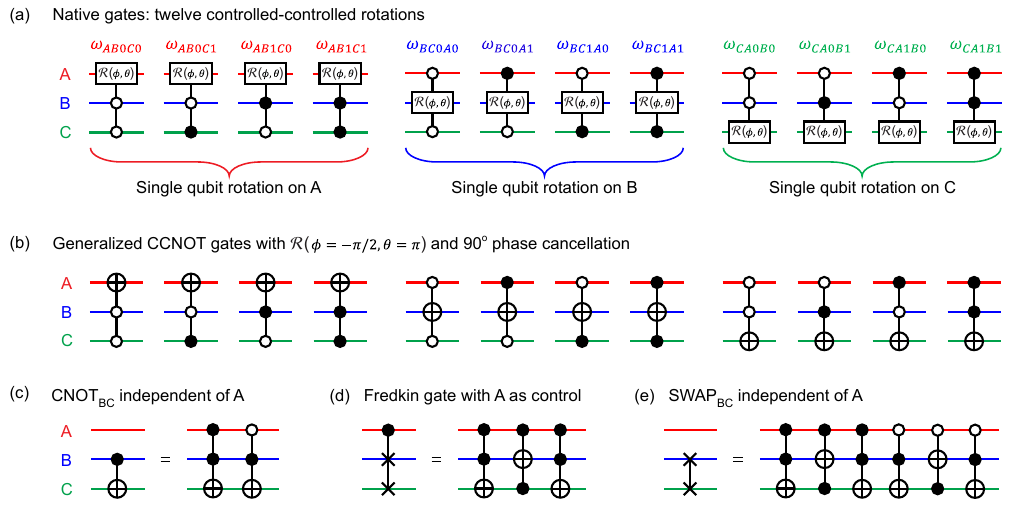}
	\caption{(a) Arbitrary rotations on each of the twelve transitions of the trimon. Applying them in sets of four as shown implements single qubit rotations on a particular qubit. (b) The generalized CCNOT gates correspond to a $\pi$-pulse on a particular transition followed by a phase adjustment as discussed in section \ref{subsec:multiqubitgates}. Examples of the decomposition of the (c) two-qubit CNOT, (d) Fredkin and (e) SWAP gates into the elementary gates available in the trimon.}
	\label{fig:qcircuits}
\end{figure*}
We now discuss single and multi-qubit operations in a trimon device. For an $N$-qubit system with all-to-all longitudinal coupling, the transition frequency of any qubit depends upon the state of the remaining $N-1$ qubits. As a result, the elementary operations in such a system are the $N2^{N-1}$ controlled-rotations, activated by the specific transition frequency. A $\pi$-pulse (anticlockwise rotation) at such a frequency implements an $(N-1)$-controlled NOT gate up to a $-90^\circ$ phase, which we call the $-i$C$_{N-1}$NOT gate. Then a true C$_{N-1}$NOT gate can be realized by appropriately adjusting the phase of all the subsequent pulses on particular transitions \cite{trimon}. We now focus on the case of a three-qubit system to understand this effect.

A three-qubit state $|ABC\rangle$ residing in an 8-dimensional Hilbert space can be expressed using the following basis vectors
\begin{equation}
|000\rangle = \begin{bmatrix}1\\0\\0\\0\\0\\0\\0\\0\end{bmatrix}, \ \
|001\rangle = \begin{bmatrix}0\\1\\0\\0\\0\\0\\0\\0\end{bmatrix}, \ \
\cdots, \ \
|111\rangle = \begin{bmatrix}0\\0\\0\\0\\0\\0\\0\\1\end{bmatrix}.
\end{equation}
We label the transition frequency of qubit C when qubits A $=|0\rangle$ and B $=|1\rangle$ as $\omega_{CA0B1}$ and similarly for others. Then the effect of a pulse at $\omega_{CA1B1}$ can be expressed as
\begin{multline}
\text{CC}\mathcal{R}_{CA1B1}(\phi,\theta) = \\
\begin{pmatrix}
1 & 0 & 0 & 0 & 0 & 0 & 0 & 0\\
0 & 1 & 0 & 0 & 0 & 0 & 0 & 0\\
0 & 0 & 1 & 0 & 0 & 0 & 0 & 0\\
0 & 0 & 0 & 1 & 0 & 0 & 0 & 0\\
0 & 0 & 0 & 0 & 1 & 0 & 0 & 0\\
0 & 0 & 0 & 0 & 0 & 1 & 0 & 0\\
0 & 0 & 0 & 0 & 0 & 0 & \cos(\theta/2) &-e^{-i\phi}\sin(\theta/2)\\
0 & 0 & 0 & 0 & 0 & 0 & e^{i\phi}\sin(\theta/2) & \cos(\theta/2)\\
\end{pmatrix},
\end{multline}
where $\theta$ is the polar angle and $\phi$ is the azimuthal angle with respect to the y-axis on the Bloch sphere for qubit C (Fig.~\ref{fig:qcircuits}(a)). Clearly, CC$\mathcal{R}_{CA1B1}(-\pi/2,\pi)$ flips the state of qubit C only when qubits A and B are in the excited state and implements an $-i$CCNOT gate on qubit C. In order to cancel this extra phase one needs to shift the phase of all subsequent pulses on qubit A with B $= |1\rangle$ and those on qubit B with A $= |1\rangle$. The phase shift needed is $-90^\circ$ ($+90^\circ$) whenever the target qubit is flipped based on the control qubit being in the $|1\rangle$ ($|0\rangle$) state. Table~\ref{table:cnot} shows shifts required for pulses at different transitions of qubits B and C after the application of various $\pi$-pulses on different transitions of qubit A. Similar rules apply for other qubits and can be extended to larger number of qubits as well. Having the CCNOT (or Toffoli) as the native gate (Fig.~\ref{fig:qcircuits}(b)) in this architecture, one can realize the Fredkin gate \cite{Fredkin1982} using three CCNOT gates (Fig.~\ref{fig:qcircuits}(d)). Similarly other gates like the two-qubit CNOT (Fig.~\ref{fig:qcircuits}(c)) and the two-qubit SWAP gate (Fig.~\ref{fig:qcircuits}(e)) can be constructed from the elementary CCNOT gates.
\begin{table}[t]
	\caption{Phase shifts required on various pulses to implement a true CCNOT gate on qubit A.}
	\begin{tabular}{c| c| c| c| c} 
		\hline
		\hline
		$\pi$-pulse at & Modify B & Shift  & Modify C & Shift \\
		\hline
		$\omega_{AB0C0}$ & $\omega_{BC0A0}, \ \omega_{BC0A1}$ & $+90^\circ$ & $\omega_{CA0B0}, \ \omega_{CA1B0}$ & $+90^\circ$ \\ 
		$\omega_{AB0C1}$ & $\omega_{BC1A0}, \ \omega_{BC1A1}$ & $+90^\circ$ & $\omega_{CA0B0}, \ \omega_{CA1B0}$ & $-90^\circ$ \\ 
		$\omega_{AB1C0}$ & $\omega_{BC0A0}, \ \omega_{BC0A1}$ & $-90^\circ$ & $\omega_{CA0B1}, \ \omega_{CA0B1}$ & $+90^\circ$ \\ 
		$\omega_{AB1C1}$ & $\omega_{BC1A0}, \ \omega_{BC1A1}$ & $-90^\circ$ & $\omega_{CA0B1}, \ \omega_{CA1B1}$ & $-90^\circ$ \\ [0ex]
		\hline
		\hline
	\end{tabular}
	\label{table:cnot}
\end{table}

Although the all-to-all coupling makes the C$_{N-1}$NOT gate very simple, single qubit rotations become less trivial, requiring application of pulses at all possible values of transition frequency for that qubit. In general, an $N$-qubit system will need pulses at 2$^{N-1}$ different frequencies for single-qubit gates, at $2^{N-2}$ frequencies for two-qubit gates, and so on. In principle it is possible to apply all these pulses simultaneously to implement a fast gate, but the process of generating and calibrating such pulses might become cumbersome beyond $N=3$. This technique of applying multi-frequency pulses is similar to an NMR technique \cite{NMR_technique_Chuang}, where a single broadband pulse covering all the frequencies is applied.

Another unique feature of such longitudinally-coupled multi-qubit system is the ability to implement error free controlled-phase gates. Let us discuss the procedure for realizing a controlled-controlled-Z (CCZ) gate for the case of a trimon. The conventional CCZ gate can be represented by
\begin{equation}
	\text{CCZ} = 
	\begin{pmatrix}
	1 & 0 & 0 & 0 & 0 & 0 & 0 & 0\\
	0 & 1 & 0 & 0 & 0 & 0 & 0 & 0\\
	0 & 0 & 1 & 0 & 0 & 0 & 0 & 0\\
	0 & 0 & 0 & 1 & 0 & 0 & 0 & 0\\
	0 & 0 & 0 & 0 & 1 & 0 & 0 & 0\\
	0 & 0 & 0 & 0 & 0 & 1 & 0 & 0\\
	0 & 0 & 0 & 0 & 0 & 0 & 1 & 0\\
	0 & 0 & 0 & 0 & 0 & 0 & 0 & -1\\
	\end{pmatrix}
	\label{eq:ccz}
\end{equation}
which flips the sign of the $|111\rangle$ state. This sign flipping can be done by simply shifting the phases of all subsequent pulses (at frequencies $\omega_{AB1C1}, \omega_{BC1A1}, \omega_{CA1B1}$) that connects $|111\rangle$ to other states by $180^\circ$. Similarly, a generalized CCZ gate which flips the sign of an arbitrary component $|ABC\rangle$ can be achieved by shifting the phases of pulses at the three relevant transition frequencies that are allowed from that particular level (see Fig.~\ref{fig:level}). Since this implementation does not involve application of a real pulse, and the microwave drive phases can be changed with high accuracy in software, the CCZ gates are calibration error free and take no time to execute\cite{ibm_zgate_free}. Two-qubit controlled-Z and single qubit Z gate then become combination of two and four CCZ gates respectively. This idea can be easily extended to impose arbitrary conditional-phase by an amount $\theta$, namely CC$\theta$ gate on any of the three-qubit components and also to larger number of longitudinally coupled qubits. Phase shifts required for realizing a CC$\theta$ gate on different basis components are tabulated in Table~\ref{table:cctheta}. Access to both generalized CCZ and CC$\theta$ gates allow significantly simpler realization of many quantum oracles. 

\begin{table}[t]
	\caption{Phase shifts required on various pulses to implement a  CC$\theta$ gate on different basis components.}
	\begin{tabular}{c| c| c| c| c| c| c} 
		\hline
		\hline
		CC$\theta$ on & Modify A & Shift  & Modify B & Shift & Modify C & Shift\\
		\hline
		$|000\rangle$ & $\omega_{AB0C0}$ & $+\theta$ & $\omega_{BC0A0}$ & +$\theta$ & $\omega_{CA0B0}$ & $+\theta$\\
		$|001\rangle$ & $\omega_{AB0C1}$ & $+\theta$ & $\omega_{BC1A0}$ & +$\theta$ & $\omega_{CA0B0}$ & $-\theta$\\
		$|010\rangle$ & $\omega_{AB1C0}$ & $+\theta$ & $\omega_{BC0A0}$ & $-\theta$ & $\omega_{CA0B1}$ & $+\theta$\\
		$|011\rangle$ & $\omega_{AB1C1}$ & $+\theta$ & $\omega_{BC1A0}$ & $-\theta$ & $\omega_{CA0B1}$ & $-\theta$\\
		$|100\rangle$ & $\omega_{AB0C0}$ & $-\theta$ & $\omega_{BC0A1}$ & +$\theta$ & $\omega_{CA1B0}$ & $+\theta$\\
		$|101\rangle$ & $\omega_{AB0C1}$ & $-\theta$ & $\omega_{BC1A1}$ & +$\theta$ & $\omega_{CA1B0}$ & $-\theta$\\
		$|110\rangle$ & $\omega_{AB1C0}$ & $-\theta$ & $\omega_{BC0A1}$ & $-\theta$ & $\omega_{CA1B1}$ & $+\theta$\\
		$|111\rangle$ & $\omega_{AB1C1}$ & $-\theta$ & $\omega_{BC1A1}$ & $-\theta$ & $\omega_{CA1B1}$ & $-\theta$
		\\ [0ex]
		\hline
		\hline
	\end{tabular}
	\label{table:cctheta}
\end{table}

\subsection{State Tomography}
\label{sec:tomo}
Performing tomography of an arbitrary $N$-qubit state requires measurement along $2^N$ basis directions. Commonly used multi-qubit architectures utilizing transverse coupling have readout resonators associated with every qubit \cite{Martinis_9xmon_nature, IBM-errordet-4qubit, riste-bit-flip, multi-qubit-3d-yale} providing the ability to measure individual qubits independently. Typical schemes used in those systems enable measurement along $\sigma_z$ direction only and pre-rotations about $x$ and $y$-axes are performed to measure along the other two orthogonal directions. Then, information from individual qubit measurements is used to compute the density matrix of the full system. On the contrary, in our system, all the qubits are (directly or indirectly) coupled to the same cavity resonator and are measured using joint dispersive readout \cite{JPA-parity, joint-readout, joint-readout-2D}, which projects the system to one of the $2^N$ energy eigenstates.
\begin{figure}[t]
	\centering
	\includegraphics[width=\columnwidth]{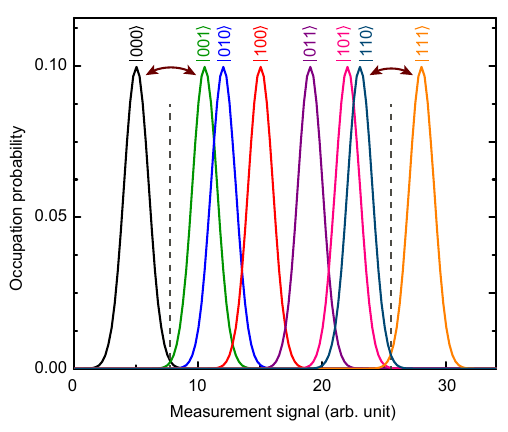}
	\caption{An example of single shot measurement histograms expected in a joint dispersive readout scheme for the eight basis states of a typical trimon device. Histograms for states $|000\rangle$ and $|111\rangle$ are quite distinguishable while those for the rest have large overlaps. Consequently, one can draw two demarcation lines to measure states $|000\rangle$ and $|111\rangle$ with high confidence and any outcome within the demarcation lines can be discarded. Then, two $\pi$ pulses at frequencies $\omega_{CA0B0}$ and $\omega_{CA1B1}$ (CCNOT gates) are applied to exchange population between pairs $|000\rangle \leftrightarrow |001\rangle$ and $|110\rangle \leftrightarrow |111\rangle$ for measurement along  $|001\rangle$ and  $|110\rangle$ as shown by the brown arrows. Next, other appropriate CCNOT gates are applied for measurement along all basis states.}
	\label{fig:dist}
\end{figure}

Representative histograms for all the basis states of a typical trimon measured using joint readout technique \cite{JPA-parity} and Josephson parametric amplifiers (JPA) \cite{JPA-Hatridge}, are shown in Fig.~\ref{fig:dist}. The overlaps between some basis states appear because dispersive shifts of the three qubits are quite similar. These overlaps make some of the distributions indistinguishable from each other (e.g., histograms for $|001\rangle$ and $|010\rangle$ are highly overlapping in Fig.~\ref{fig:dist}). However, this problem can be easily overcome by noting that states $|000\rangle$ and $|111\rangle$ have extremely small overlap with the rest and can be measured with high confidence. Then one can draw two demarcation lines (gray lines in Fig.~\ref{fig:dist}) to separate states $|000\rangle$ and $|111\rangle$ from the rest, and discard any outcomes which fall between the two lines. Thus, in the first measurement, one finds projections along $|000\rangle$ and $|111\rangle$. Then to find projections along $|001\rangle$ and $|110\rangle$ two CCNOT gates are applied at frequencies $\omega_{CA0B0}$ and $\omega_{CA1B1}$ to exchange population between pairs $|000\rangle \leftrightarrow |001\rangle$ and $|110\rangle \leftrightarrow |111\rangle$. In the next iteration two more CCNOT gates are applied to perform measurements along  $|010\rangle$ and $|101\rangle$ and so on. In the case of a trimon one needs four rounds of measurements to find projections of all three-qubits along $\sigma_z$ direction. Then this whole process has to be repeated with all combinations of pre-rotations of individual qubits along $x$ or $y$-axes to find all necessary projections to reconstruct the density matrix. This technique can be extended to systems with larger number of qubits and truncated to perform tomography of a smaller subspace of the full Hilbert space\cite{trimon}.

Measurement errors in this technique will have a contribution from overlap of population distributions on either side of the demarcation lines shown in Fig.~\ref{fig:dist}. This in turn depends on usual cQED readout parameters\cite{trimon} like dispersive shifts, measurement power, integration times and system noise temperature of the amplification chain. Any error in the implementation of the CCNOT gates (usually small; see section~\ref{subsec:sim}) in the various steps of state tomography will further add to the overall measurement error.

\subsection{Device Parameter Optimization}
\label{subsec:dev_par}

We now discuss how to optimize the device parameters for a trimon to enable efficient three qubit operation. This involves finding the Josephson energies of the four junctions and all the inter-node capacitors and translate that to a real device design. As mentioned in Section \ref{subsec:levels}, the main optimization is to ensure that all the twelve transitions in the computational subspace are spectroscopically distinct so that each transition can be separately addressed and with sufficient speed. In addition, these transitions should also be spectroscopically distinct from $|1\rangle \rightarrow |2\rangle$ transition for each mode to prevent leakage out of the computational subspace. This implies that the self-Kerr ($J_A,J_B,J_C$) and the cross-Kerr ($J_{AB}, J_{BC}, J_{CA}$) shifts should all be sufficiently distinct from each other. Another constraint we impose is that all transitions should lie roughly in the $4-6$ GHz range so that they are sufficiently detuned from our measurement cavity ($\sim7.3$ GHz) to suppress Purcell decay.

\setlength{\tabcolsep}{1.5 mm}
\begin{table*}[t]
	\caption{A set of design parameters (Josephson energies and capacitances) and corresponding device parameters (frequencies, anharmonicites, inter-qubit coupling strengths, qubit-cavity coupling strengths and dispersive shifts) for optimal performance.}
	\begin{tabular}{||c c| c c| c c| c c| c c| c c| c c| c c||} 
		\hline
		\hline
		$E_J$ & (GHz) & $C$  & (fF) & $C$ & (fF) & Freq. & (GHz) & $\alpha$ & (GHz) & Coupling & (MHz) & $g'$ & (MHz) & $\chi$ & (MHz) \\
		\hline
		$E_{J_{12}}$ & 8.794 & $C_{12}$ & 34.0 & $C_{11}$ & 0.01 & $f_A$ & 5.244 & $\alpha_A$ & $-0.120$ & $J_{AB}$ & 81 & $g'_A$ & 69 & $\chi_A$ & 0.131\\
		$E_{J_{23}}$ & 8.712 & $C_{23}$ & 34.0 & $C_{22}$ & 0.02 & $f_B$ & 4.773 & $\alpha_B$ & $-0.114$ & $J_{BC}$ & 99 & $g'_B$ & 13 & $\chi_B$ & 0.089\\
		$E_{J_{34}}$ & 8.042 & $C_{34}$ & 34.0 & $C_{33}$ & 0.01 & $f_C$ & 6.059 & $\alpha_C$ & $-0.151$ & $J_{CA}$ & 117 & $g'_C$ & 5 & $\chi_C$ & 0.123\\
		$E_{J_{41}}$ & 7.143 & $C_{41}$ & 34.0 & $C_{44}$ & 0.02 &  &   &   &   &   &   &   &   &   &  \\
		  &   & $C_{13}$ & 11.2 &   &   &   &   &   &   &   &   &   &   &   &  \\
		  &   & $C_{24}$ & 19.1 &   &   &   &   &   &   &   &   &   &   &   &  
		\\ [0ex]
		\hline
		\hline
	\end{tabular}
	\label{table:sim}
\end{table*}

We first developed a numerical code which outputs all trimon parameters given the junction and capacitor values. To find a device with optimal parameters we start with a target level spacing based on the constraints discussed above and run a minimization routine to arrive at the appropriate junction energies and capacitances that provides the level spacings closest to the target values. One can choose whether to introduce asymmetry in the junctions, capacitors or both in the optimization process. The qubit-cavity coupling and dispersive shifts are calculated by using $g_A\sim70$ MHz which is typical for transmons in circuit QED geometry. The process is iterated with minor adjustments to the average Josephson energy which scales all mode frequencies, till all transitions are separated by about 30 MHz or higher and dispersive shifts for each mode are large enough to achieve measurement histogram separation (Section \ref{sec:tomo}). The introduction of asymmetry in the structure also provides finite coupling of all three modes to the cavity which enables exciting all modes with reasonable microwave power and avoids complications due to ac Stark shifts of the modes due to the global microwave drive.

Once the device parameters are obtained, a finite element simulation is performed to find the 3D capacitor geometry that would give rise to the required capacitance matrix. The Josephson energies are converted to junction areas from fabrication calibrations. In Table~\ref{table:sim} we provide one such design and predicted parameters where asymmetry is allowed only in the Josephson junctions. This design provides frequency separation larger than 35 MHz between any pairs (including $|1\rangle \rightarrow |2\rangle$ transitions).

The native gate operations in this device are the controlled-controlled-rotations. Implementing and calibrating these rotations is identical to calibrating single qubit gates in a standard qubit. The fidelity of these gates is then predominantly determined by the decoherence time and the gate speed. Just like in a standard transmon qubit, the gate speed is then restricted by how short a pulse one can generate without exciting neighboring transitions. As demonstrated earlier \cite{trimon}, gate fidelity of $\sim0.99$ is achievable and can be further improved by optimizing pulse shapes and level spacing. Single and two qubit gates are composed of several such elementary gates and the overall fidelity will just depend on the total number of elementary gates used in a particular operation. Further, the multiple transitions to be addressed for two-qubit and single qubit gates can be achieved by a multi-frequency pulse which can further improve fidelity by reducing gate time. The only caveat is that none of the transitions should involve a common level.

\subsection{Numerical estimation of state/gate fidelity}
\label{subsec:sim}

\begin{figure}[t]
	\centering
	\includegraphics[width=\columnwidth]{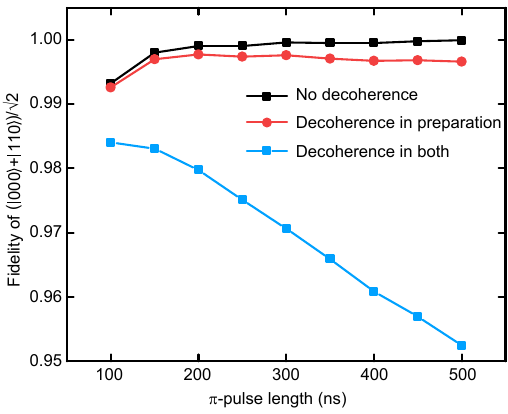}
	\caption{Fidelity of the Bell state $(|000\rangle + |110\rangle)/\sqrt{2}$ as a function of $\pi$-pulse length used for state preparation. The $pi/2$-pulses have lengths half of that of the $\pi$-pulses. The black curve represents simulations in the absence of any decoherence, relaxation error is introduced for the red curve and the blue curve shows fidelities when relaxation is present in both state preparation and tomography pulses.}
	\label{fig:fidelity}
\end{figure}

In order to estimate the performance of our device, we performed numerical simulations using the QuTiP open-source software\cite{qutip1,qutip2}. We simulated the Hamiltonian given by Eq.~\eqref{eq:Hsys} (neglecting the three-body term) for a trimon system considering up to three levels of each mode to include the effect of leakage out of the computational subspace. We used square-envelop pulses and varied the length of the $\pi$ pulses (kept same for all the twelve transitions for a particular numerical experiment) by adjusting the drive amplitudes. The $\pi/2$-pulse length is then just half of that value. The fidelity of the Bell state $(|000\rangle + |110\rangle)/\sqrt{2}$ as a function of pulse length is shown in Fig.~\ref{fig:fidelity} (black curve) where the tomography is performed by following the protocol discussed in section~\ref{sec:tomo} without considering any decoherence and measurement error. We used the standard definition of fidelity $\mathcal{F} = \text{Tr}\left[ \sqrt{\sqrt{\rho_{\rm id}} \rho_{\rm MLE} \sqrt{\rho_{\rm id}}} \right]$, where $\rho_{\rm id}$ is the ideal density matrix and $\rho_{\rm MLE}$ corresponds to the density matrix obtained from maximum likelihood estimation\cite{MLE, MLE1}. The reduced fidelity for shorter pulses is a result of information leakage, while use of longer pulses (having reduced bandwidth) does not improve the fidelity significantly. In order to mimic the performance of a realistic device, we introduced relaxations with T$_1$ values of 50 $\mu$s, 40 $\mu$s and 30 $\mu$s respectively for qubits A, B and C. We first included the effect of relaxation in the state preparation step only to determine the intrinsic fidelity and the results are depicted in Fig.~\ref{fig:fidelity} (red curve). The fidelity slowly decreases with longer pulses. The blue curve represents the same result when decoherence is also included in the tomography and shows much faster fall of fidelity with increasing pulse lengths as the tomography involves application of a large number of pulses.
	
From Fig.~\ref{fig:fidelity}, we conclude 200 ns to be the optimal length for the $\pi$-pulses and simulated single transition randomized benchmarking\cite{Chow-RB-PRL} (RB) to determine the average fidelities of the CC$\mathcal{R}$ gates (gates that apply $\pi$ and $\pi/2$ rotations). Note that each RB simulation essentially involves performing Rabi rotations between the two energy eigenstates connected by the particular transition being addressed and thus is expected to be of very high fidelity. As examples, the average gate fidelities for qubit A were found to be 0.998(2), 0.995(3), 0.996(3) and 0.993(2) for the transitions at $\omega_{AB0C0}, \ \omega_{AB0C1}, \ \omega_{AB1C0}$ and $\omega_{AB1C1}$ respectively. As a measure of the performance of the trimon device, we prepared various three-qubit states with high fidelities as shown in Table~\ref{table:sim2}. The fidelities can further be improved by using shorter pulses with appropriate pulse shaping\cite{drag-pulse}. As mentioned earlier, the fidelity for single and two-qubit gates which involve multiple transitions can be calculated by appropriately combining the individual transition fidelities. These can be further improved by using multi-frequency pulses as explained earlier. Clearly, the estimated performance of the trimon makes it an excellent candidate for being used as a high-fidelity three-qubit building block for a larger multi-qubit system.

\begin{table}[t]
	\centering
	\caption{Simulated fidelities $\mathcal{F}$ of various important three-qubit states. $\pi$-pulse lengths of 200 ns were used for all transitions. The second column shows intrinsic fidelities where decoherence is considered only in the preparation pulses. The third column displays the same with decoherence introduced in the tomographic pulses as well.}
	\label{table:sim2}
	\begin{tabular}{|c||c|c|}
		\hline
		\multirow{2}{*}{State} 
		& \multicolumn{2}{c|}{$\mathcal{F}$ with decoherence in} \\             \cline{2-3}
		& Prep. only & Prep. \& tomography \\  \hline
		$\dfrac{|000\rangle + |110\rangle}{\sqrt{2}}$ & 0.9977 & 0.9798 \\ [3 ex]
		$\dfrac{|000\rangle + |011\rangle}{\sqrt{2}}$ & 0.9962 & 0.9777 \\ [3 ex]
		$\dfrac{|000\rangle + |101\rangle}{\sqrt{2}}$ & 0.9963 & 0.9754 \\ [3 ex]
		$\dfrac{|000\rangle + |111\rangle}{\sqrt{2}}$ & 0.9938 & 0.9695 \\ [3 ex]
		$\dfrac{|001\rangle + |010\rangle + |100\rangle}{\sqrt{3}}$ & 0.9949 & 0.9737  \\
		[3 ex]
		$\dfrac{|011\rangle + |110\rangle + |101\rangle}{\sqrt{3}}$ & 0.9839 & 0.9653  \\
		[3 ex]
		$\dfrac{(|0\rangle + |1\rangle)^{\otimes 3}}{2\sqrt{2}}$ & 0.9876 & 0.9699 \\ \hline
	\end{tabular}
\end{table}


\subsection{Coupling multiple trimons}
\label{subsec:mult_trimon}

\begin{figure}[t]
	\centering
	\includegraphics[width=\columnwidth]{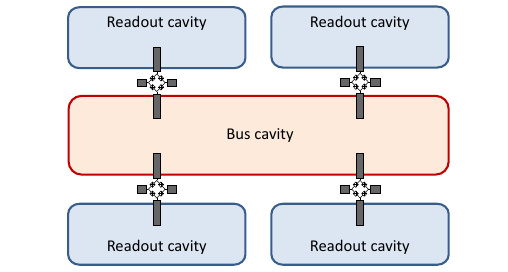}
	\caption[Possible design for building larger processor]{A possible design for scaling up to a 12-qubit processor in 3D architecture using four trimons. Each trimon has its own readout cavity while a common bus-cavity provides coupling between them.}
	\label{fig:trimon_bus}
\end{figure}

In order to build a processor with larger number of qubits, we can adapt the architecture for coupling different transmons\cite{wallraff-dig-qsim, IBM-errordet-4qubit, riste-bit-flip} where multiple trimon blocks having their own readout resonators are coupled to each other via bus resonators. One can also use a common bus resonator as shown in Fig.~\ref{fig:trimon_bus} and used to demonstrate the resonator-induced phase (RIP) gate in a multi-qubit 3D cQED system\cite{RIP-gate}. The intra-trimon gates would be realized by applying microwave pulses through individual readout cavities while gates between different trimons could be implemented by adapting well-established techniques like cross-resonance\cite{CR-gate}. Note that only one of the trimon modes is to be coupled to the bus resonator as any uncoupled mode (to the bus) can always be swapped with the coupled one. One can also implement a frequency-multiplexed readout scheme\cite{multiplex-martinis} for performing simultaneous measurement of the individual cavities where a broadband parametric amplifier\cite{impa, TWPA-Berkeley, TWPA-martinis,BBparamp} will prove to be useful in minimizing the resources.

\section{Other MULTIMON Geometries}

\begin{figure}[t]
	\centering
	\includegraphics[width=\columnwidth]{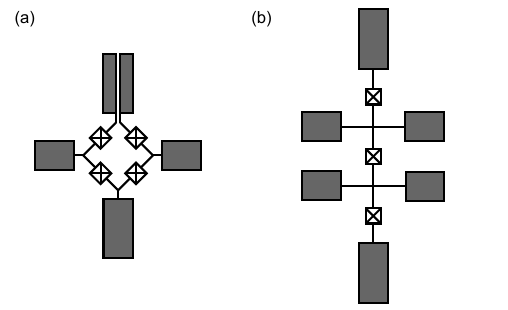}
	\caption{Schematic of multimon devices in (a) open ring, and (b) linear chain configuration in the 3D geometry (not to scale). Open ring design is achieved by splitting one of the capacitors of a ring multimon to eliminate the flux degree of freedom. This splitting introduces an extra low frequency mode with small anharmonicity and can be ignored by leaving it in its ground state. A linear chain is topologically equivalent to the open ring but with a different capacitance matrix and usually leads to larger anharmonicities and inter-qubit coupling.}
	\label{fig:designs}
\end{figure}

The technique described in section \ref{sec:formalism} is completely general and can be applied to any device geometry, characterized by different inductive energy and capacitance matrices, provided one remains in the weakly anharmonic oscillator limit. In this section we mention a few other prospective designs. The first one is the open ring multimon device, which can be built by simply splitting one of the capacitor pads as shown in Fig.~\ref{fig:designs}(a) (in 3D geometry). This leads to a new mode whose frequency and anharmonicity can be made small by using a large capacitance between the split node. This almost linear low frequency mode can be ignored during experiments by leaving it in its ground state.

Another promising design is the linear chain (Fig.~\ref{fig:designs}(b)) which can be obtained by unwrapping the split ring design. Now for $N$ qubit modes, we need only $N$ Josephson junctions. The modes of a linear chain are in general non-degenerate and only become degenerate in the case of zero capacitance between non-nearest nodes with identical junctions and nearest-neighbor capacitances. In general, this geometry provides larger anharmonicity and coupling strengths as compared to ring geometry. A general property of designs with no loop is that multi-body transverse coupling terms ($\tilde{\Phi}_\mu \tilde{\Phi}_\nu \cdots \tilde{\Phi}_\zeta$) are never present in the system.

\section{Conclusions}

In this article, we introduced the concept of using multi-mode superconducting circuits to construct a system of multiple, strongly coupled transmon-like qubits, nicknamed ``multimon". We described a general method to analyze such circuits and showed that it leads to longitudinal coupling between each pair of qubits. We presented a detailed analysis and discussed properties of the ``ring" type multimon devices where the Josephson junctions are incorporated in a single loop structure. We then focused on the three-qubit version called the trimon and discussed how to extract and optimize all relevant parameters, perform gate operations, and implement quantum state tomography to build a practical three-qubit quantum processor. An essential feature of the trimon is the single-pulse universal CCNOT gate which can be implemented with high fidelity. Further, access to error-free generalized CCZ gates can help in simpler realization of many quantum algorithms. While multimon devices realizing more than three qubits can offer excellent inter-qubit connectivity, it might be impractical to generate the control pulses and perform joint dispersive readout to be of use as an efficient quantum processor. We envisage the construction of larger processors using several trimon blocks coupled via a common bus resonator, and adapting cross-resonance \cite{CR-gate} and resonance-induced-phase \cite{RIP-gate} gates to implement inter-trimon quantum operations. The all-to-all longitudinal coupling in multimons might find applications in quantum annealing as well \cite{Transmon-annealer-Zoller}. Further, the protected modes of the trimon can also be used as an effective single qubit which can be optimized for strong measurability without sacrificing coherence \cite{Auffeves_Buisson-theory}. Finally, the dissimilar coupling of the three modes of the trimon to its environment\cite{trimon} (cavity, qubit or bath) can be exploited to implement switchable coupling schemes with all-microwave control.

\textit{Acknowledgements:} This work was supported by the Department of Atomic Energy of Government of India. R.V. acknowledges funding from the Department of Science and Technology, India via the Ramanujan Fellowship.

\appendix
\section{Dispersive shifts}
\label{app:dispersive}
In the most general case, each of the three qubits of a trimon will have both direct and indirect coupling. The resulting dispersive shifts of the seven eigenstates are given by,
\begin{widetext}
	\begin{subequations}
		\label{eq:chi_full}
		\begin{align}
		\chi_A = \chi_{100} = \ &g_A'^2 \left( \dfrac{1}{\Delta_{A0}} - \dfrac{1}{\Delta_{A0} - 2J_A} \right) + \dfrac{g_B'^2}{2} \left( \dfrac{1}{\Delta_{B0}} - \dfrac{1}{\Delta_{B0}-2J_{AB}} \right) + \dfrac{g_C'^2}{2} \left( \dfrac{1}{\Delta_{C0}} - \dfrac{1}{\Delta_{C0}-2J_{CA}} \right),\\[3pt]
		\chi_B = \chi_{010} = \ &\dfrac{g_A'^2}{2} \left( \dfrac{1}{\Delta_{A0}} - \dfrac{1}{\Delta_{A0}-2J_{AB}} \right) + g_B'^2 \left( \dfrac{1}{\Delta_{B0}} - \dfrac{1}{\Delta_{B0}-2J_{B}} \right) + \dfrac{g_C'^2}{2} \left( \dfrac{1}{\Delta_{C0}} - \dfrac{1}{\Delta_{C0}-2J_{BC}} \right),\\[3pt]
		\chi_C = \chi_{001} = \ &\dfrac{g_A^2}{2} \left( \dfrac{1}{\Delta_{A0}} - \dfrac{1}{\Delta_{A0}-2J_{CA}} \right) + \dfrac{g_B'^2}{2} \left( \dfrac{1}{\Delta_{B0}} - \dfrac{1}{\Delta_{B0}-2J_{BC}} \right) + g_C'^2 \left( \dfrac{1}{\Delta_{C0}} - \dfrac{1}{\Delta_{C0}-2J_{C}} \right),\\
		\begin{split} 
		\chi_{AB} = \chi_{110} = \ &\dfrac{g_A'^2}{2} \bigg( \dfrac{1}{\Delta_{A0}} + \dfrac{1}{\Delta_{A0} - 2J_{AB}} - \dfrac{2}{\Delta_{A0} - 2J_A - 2J_{AB}} \bigg)\\
		+ &\dfrac{g_B'^2}{2} \left( \dfrac{1}{\Delta_{B0}} + \dfrac{1}{\Delta_{B0}-2J_{AB}} - \dfrac{2}{\Delta_{B0}-2J_B-2J_{AB}} \right)
		+ \dfrac{g_C'^2}{2} \left( \dfrac{1}{\Delta_{C0}} - \dfrac{1}{\Delta_{C0}-2J_{CA}-2J_{BC}} \right),
		\end{split}\\	
		\begin{split} 
		\chi_{BC} = \chi_{011} = \ &\dfrac{g_A'^2}{2} \left( \dfrac{1}{\Delta_{A0}} - \dfrac{1}{\Delta_{A0} - 2J_{AB} - 2J_{CA}} \right)
		+ \dfrac{g_B'^2}{2} \left( \dfrac{1}{\Delta_{B0}} + \dfrac{1}{\Delta_{B0}-2J_{BC}} - \dfrac{2}{\Delta_{B0}-2J_B-2J_{BC}} \right) \\[3pt]
		+ &\dfrac{g_C'^2}{2} \left( \dfrac{1}{\Delta_{C0}} +\dfrac{1}{\Delta_{C0}-2J_{BC}} - \dfrac{2}{\Delta_{C0}-2J_{C}-2J_{BC}} \right),
		\end{split}\\
		\begin{split} 
		\chi_{CA} = \chi_{101} = \ &\dfrac{g_A'^2}{2} \left( \dfrac{1}{\Delta_{A0}} +\dfrac{1}{\Delta_{A0}-2J_{CA}} - \dfrac{2}{\Delta_{A0} - 2J_{A} - 2J_{CA}} \right)
		+ \dfrac{g_B'^2}{2} \left( \dfrac{1}{\Delta_{B0}} - \dfrac{1}{\Delta_{B0}-2J_{BC}-2J_{AB}} \right) \\[3pt]
		+ &\dfrac{g_C'^2}{2} \left( \dfrac{1}{\Delta_{C0}} +\dfrac{1}{\Delta_{C0}-2J_{CA}} - \dfrac{2}{\Delta_{C0}-2J_{C}-2J_{CA}} \right),
		\end{split}\\
		\begin{split}
		\chi_{ABC} = \chi_{111} = \ &\dfrac{g_A'^2}{2} \left( \dfrac{1}{\Delta_{A0}} +\dfrac{1}{\Delta_{A0}-2J_{AB}-2J_{CA}} - \dfrac{2}{\Delta_{A0} - 2J_{A}-2J_{AB} - 2J_{CA}} \right)\\[3pt]
		+ &\dfrac{g_B'^2}{2} \left( \dfrac{1}{\Delta_{B0}} + \dfrac{1}{\Delta_{B0}-2J_{BC}-2J_{AB}} - \dfrac{2}{\Delta_{B0}-2J_B-2J_{BC}-2J_{AB}} \right) \\[3pt]
		+ &\dfrac{g_C'^2}{2} \left( \dfrac{1}{\Delta_{C0}} +\dfrac{1}{\Delta_{C0}-2J_{CA}-2J_{BC}} - \dfrac{2}{\Delta_{C0}-2J_{C}-2J_{CA}-2J_{BC}} \right),
		\end{split}
		\end{align}
	\end{subequations}
\end{widetext}
with 
\begin{subequations}
	\begin{align}
	\Delta_{A0} &=\omega_A-2J_A-2J_{AB}-2J_{CA}-\omega_{r},\\
	\Delta_{B0} &=\omega_B-2J_B-2J_{BC}-2J_{AB}-\omega_{r},\\
	\Delta_{C0} &=\omega_C-2J_C-2J_{CA}-2J_{BC}-\omega_{r}.		
	\end{align}
\end{subequations}


\begin{thebibliography}{58}%
	\makeatletter
	\providecommand \@ifxundefined [1]{%
		\@ifx{#1\undefined}
	}%
	\providecommand \@ifnum [1]{%
		\ifnum #1\expandafter \@firstoftwo
		\else \expandafter \@secondoftwo
		\fi
	}%
	\providecommand \@ifx [1]{%
		\ifx #1\expandafter \@firstoftwo
		\else \expandafter \@secondoftwo
		\fi
	}%
	\providecommand \natexlab [1]{#1}%
	\providecommand \enquote  [1]{``#1''}%
	\providecommand \bibnamefont  [1]{#1}%
	\providecommand \bibfnamefont [1]{#1}%
	\providecommand \citenamefont [1]{#1}%
	\providecommand \href@noop [0]{\@secondoftwo}%
	\providecommand \href [0]{\begingroup \@sanitize@url \@href}%
	\providecommand \@href[1]{\@@startlink{#1}\@@href}%
	\providecommand \@@href[1]{\endgroup#1\@@endlink}%
	\providecommand \@sanitize@url [0]{\catcode `\\12\catcode `\$12\catcode
		`\&12\catcode `\#12\catcode `\^12\catcode `\_12\catcode `\%12\relax}%
	\providecommand \@@startlink[1]{}%
	\providecommand \@@endlink[0]{}%
	\providecommand \url  [0]{\begingroup\@sanitize@url \@url }%
	\providecommand \@url [1]{\endgroup\@href {#1}{\urlprefix }}%
	\providecommand \urlprefix  [0]{URL }%
	\providecommand \Eprint [0]{\href }%
	\providecommand \doibase [0]{http://dx.doi.org/}%
	\providecommand \selectlanguage [0]{\@gobble}%
	\providecommand \bibinfo  [0]{\@secondoftwo}%
	\providecommand \bibfield  [0]{\@secondoftwo}%
	\providecommand \translation [1]{[#1]}%
	\providecommand \BibitemOpen [0]{}%
	\providecommand \bibitemStop [0]{}%
	\providecommand \bibitemNoStop [0]{.\EOS\space}%
	\providecommand \EOS [0]{\spacefactor3000\relax}%
	\providecommand \BibitemShut  [1]{\csname bibitem#1\endcsname}%
	\let\auto@bib@innerbib\@empty
	\bibitem [{\citenamefont {Devoret}\ and\ \citenamefont
		{Schoelkopf}(2013)}]{sup-qubit-review-science}%
	\BibitemOpen
	\bibfield  {author} {\bibinfo {author} {\bibfnamefont {M.~H.}\ \bibnamefont
			{Devoret}}\ and\ \bibinfo {author} {\bibfnamefont {R.~J.}\ \bibnamefont
			{Schoelkopf}},\ }\href {\doibase 10.1126/science.1231930} {\bibfield
		{journal} {\bibinfo  {journal} {Science}\ }\textbf {\bibinfo {volume}
			{339}},\ \bibinfo {pages} {1169} (\bibinfo {year} {2013})}\BibitemShut
	{NoStop}%
	\bibitem [{\citenamefont {You}\ and\ \citenamefont {Nori}(2011)}]{qoptics_rev}%
	\BibitemOpen
	\bibfield  {author} {\bibinfo {author} {\bibfnamefont {J.~Q.}\ \bibnamefont
			{You}}\ and\ \bibinfo {author} {\bibfnamefont {F.}~\bibnamefont {Nori}},\
	}\href {\doibase 10.1038/nature10122} {\bibfield  {journal} {\bibinfo
			{journal} {Nature}\ }\textbf {\bibinfo {volume} {474}},\ \bibinfo {pages}
		{589} (\bibinfo {year} {2011})}\BibitemShut {NoStop}%
	\bibitem [{\citenamefont {Roy}\ and\ \citenamefont
		{Devoret}(2016)}]{paramp_review}%
	\BibitemOpen
	\bibfield  {author} {\bibinfo {author} {\bibfnamefont {A.}~\bibnamefont
			{Roy}}\ and\ \bibinfo {author} {\bibfnamefont {M.}~\bibnamefont {Devoret}},\
	}\href {\doibase https://doi.org/10.1016/j.crhy.2016.07.012} {\bibfield
		{journal} {\bibinfo  {journal} {Comptes Rendus Physique}\ }\textbf {\bibinfo
			{volume} {17}},\ \bibinfo {pages} {740 } (\bibinfo {year} {2016})},\ \bibinfo
	{note} {quantum microwaves / Micro-ondes quantiques}\BibitemShut {NoStop}%
	\bibitem [{\citenamefont {Nation}\ \emph {et~al.}(2012)\citenamefont {Nation},
		\citenamefont {Johansson}, \citenamefont {Blencowe},\ and\ \citenamefont
		{Nori}}]{amplifying_vacuum_review}%
	\BibitemOpen
	\bibfield  {author} {\bibinfo {author} {\bibfnamefont {P.~D.}\ \bibnamefont
			{Nation}}, \bibinfo {author} {\bibfnamefont {J.~R.}\ \bibnamefont
			{Johansson}}, \bibinfo {author} {\bibfnamefont {M.~P.}\ \bibnamefont
			{Blencowe}}, \ and\ \bibinfo {author} {\bibfnamefont {F.}~\bibnamefont
			{Nori}},\ }\href {\doibase 10.1103/RevModPhys.84.1} {\bibfield  {journal}
		{\bibinfo  {journal} {Rev. Mod. Phys.}\ }\textbf {\bibinfo {volume} {84}},\
		\bibinfo {pages} {1} (\bibinfo {year} {2012})}\BibitemShut {NoStop}%
	\bibitem [{\citenamefont {Xiang}\ \emph {et~al.}(2013)\citenamefont {Xiang},
		\citenamefont {Ashhab}, \citenamefont {You},\ and\ \citenamefont
		{Nori}}]{hybrid_sc}%
	\BibitemOpen
	\bibfield  {author} {\bibinfo {author} {\bibfnamefont {Z.-L.}\ \bibnamefont
			{Xiang}}, \bibinfo {author} {\bibfnamefont {S.}~\bibnamefont {Ashhab}},
		\bibinfo {author} {\bibfnamefont {J.~Q.}\ \bibnamefont {You}}, \ and\
		\bibinfo {author} {\bibfnamefont {F.}~\bibnamefont {Nori}},\ }\href {\doibase
		10.1103/RevModPhys.85.623} {\bibfield  {journal} {\bibinfo  {journal} {Rev.
				Mod. Phys.}\ }\textbf {\bibinfo {volume} {85}},\ \bibinfo {pages} {623}
		(\bibinfo {year} {2013})}\BibitemShut {NoStop}%
	\bibitem [{\citenamefont {Nakamura}\ \emph {et~al.}(1999)\citenamefont
		{Nakamura}, \citenamefont {Pashkin},\ and\ \citenamefont
		{Tsai}}]{nakamura_cpb}%
	\BibitemOpen
	\bibfield  {author} {\bibinfo {author} {\bibfnamefont {Y.}~\bibnamefont
			{Nakamura}}, \bibinfo {author} {\bibfnamefont {Y.~A.}\ \bibnamefont
			{Pashkin}}, \ and\ \bibinfo {author} {\bibfnamefont {J.}~\bibnamefont
			{Tsai}},\ }\href@noop {} {\bibfield  {journal} {\bibinfo  {journal} {Nature}\
		}\textbf {\bibinfo {volume} {398}},\ \bibinfo {pages} {786} (\bibinfo {year}
		{1999})}\BibitemShut {NoStop}%
	\bibitem [{\citenamefont {Lin}\ \emph {et~al.}(2018)\citenamefont {Lin},
		\citenamefont {Nguyen}, \citenamefont {Grabon}, \citenamefont {San~Miguel},
		\citenamefont {Pankratova},\ and\ \citenamefont
		{Manucharyan}}]{fluxonium_ms_manucharyan}%
	\BibitemOpen
	\bibfield  {author} {\bibinfo {author} {\bibfnamefont {Y.-H.}\ \bibnamefont
			{Lin}}, \bibinfo {author} {\bibfnamefont {L.~B.}\ \bibnamefont {Nguyen}},
		\bibinfo {author} {\bibfnamefont {N.}~\bibnamefont {Grabon}}, \bibinfo
		{author} {\bibfnamefont {J.}~\bibnamefont {San~Miguel}}, \bibinfo {author}
		{\bibfnamefont {N.}~\bibnamefont {Pankratova}}, \ and\ \bibinfo {author}
		{\bibfnamefont {V.~E.}\ \bibnamefont {Manucharyan}},\ }\href {\doibase
		10.1103/PhysRevLett.120.150503} {\bibfield  {journal} {\bibinfo  {journal}
			{Phys. Rev. Lett.}\ }\textbf {\bibinfo {volume} {120}},\ \bibinfo {pages}
		{150503} (\bibinfo {year} {2018})}\BibitemShut {NoStop}%
	\bibitem [{\citenamefont {Earnest}\ \emph {et~al.}(2018)\citenamefont
		{Earnest}, \citenamefont {Chakram}, \citenamefont {Lu}, \citenamefont
		{Irons}, \citenamefont {Naik}, \citenamefont {Leung}, \citenamefont {Ocola},
		\citenamefont {Czaplewski}, \citenamefont {Baker}, \citenamefont {Lawrence},
		\citenamefont {Koch},\ and\ \citenamefont
		{Schuster}}]{fluxonium_ms_schuster}%
	\BibitemOpen
	\bibfield  {author} {\bibinfo {author} {\bibfnamefont {N.}~\bibnamefont
			{Earnest}}, \bibinfo {author} {\bibfnamefont {S.}~\bibnamefont {Chakram}},
		\bibinfo {author} {\bibfnamefont {Y.}~\bibnamefont {Lu}}, \bibinfo {author}
		{\bibfnamefont {N.}~\bibnamefont {Irons}}, \bibinfo {author} {\bibfnamefont
			{R.~K.}\ \bibnamefont {Naik}}, \bibinfo {author} {\bibfnamefont
			{N.}~\bibnamefont {Leung}}, \bibinfo {author} {\bibfnamefont
			{L.}~\bibnamefont {Ocola}}, \bibinfo {author} {\bibfnamefont {D.~A.}\
			\bibnamefont {Czaplewski}}, \bibinfo {author} {\bibfnamefont
			{B.}~\bibnamefont {Baker}}, \bibinfo {author} {\bibfnamefont
			{J.}~\bibnamefont {Lawrence}}, \bibinfo {author} {\bibfnamefont
			{J.}~\bibnamefont {Koch}}, \ and\ \bibinfo {author} {\bibfnamefont {D.~I.}\
			\bibnamefont {Schuster}},\ }\href {\doibase 10.1103/PhysRevLett.120.150504}
	{\bibfield  {journal} {\bibinfo  {journal} {Phys. Rev. Lett.}\ }\textbf
		{\bibinfo {volume} {120}},\ \bibinfo {pages} {150504} (\bibinfo {year}
		{2018})}\BibitemShut {NoStop}%
	\bibitem [{\citenamefont {You}\ \emph {et~al.}(2007)\citenamefont {You},
		\citenamefont {Hu}, \citenamefont {Ashhab},\ and\ \citenamefont
		{Nori}}]{shunted-flux-qubit}%
	\BibitemOpen
	\bibfield  {author} {\bibinfo {author} {\bibfnamefont {J.~Q.}\ \bibnamefont
			{You}}, \bibinfo {author} {\bibfnamefont {X.}~\bibnamefont {Hu}}, \bibinfo
		{author} {\bibfnamefont {S.}~\bibnamefont {Ashhab}}, \ and\ \bibinfo {author}
		{\bibfnamefont {F.}~\bibnamefont {Nori}},\ }\href {\doibase
		10.1103/PhysRevB.75.140515} {\bibfield  {journal} {\bibinfo  {journal} {Phys.
				Rev. B}\ }\textbf {\bibinfo {volume} {75}},\ \bibinfo {pages} {140515}
		(\bibinfo {year} {2007})}\BibitemShut {NoStop}%
	\bibitem [{\citenamefont {Koch}\ \emph {et~al.}(2007)\citenamefont {Koch},
		\citenamefont {Yu}, \citenamefont {Gambetta}, \citenamefont {Houck},
		\citenamefont {Schuster}, \citenamefont {Majer}, \citenamefont {Blais},
		\citenamefont {Devoret}, \citenamefont {Girvin},\ and\ \citenamefont
		{Schoelkopf}}]{transmon_theory}%
	\BibitemOpen
	\bibfield  {author} {\bibinfo {author} {\bibfnamefont {J.}~\bibnamefont
			{Koch}}, \bibinfo {author} {\bibfnamefont {T.~M.}\ \bibnamefont {Yu}},
		\bibinfo {author} {\bibfnamefont {J.}~\bibnamefont {Gambetta}}, \bibinfo
		{author} {\bibfnamefont {A.~A.}\ \bibnamefont {Houck}}, \bibinfo {author}
		{\bibfnamefont {D.~I.}\ \bibnamefont {Schuster}}, \bibinfo {author}
		{\bibfnamefont {J.}~\bibnamefont {Majer}}, \bibinfo {author} {\bibfnamefont
			{A.}~\bibnamefont {Blais}}, \bibinfo {author} {\bibfnamefont {M.~H.}\
			\bibnamefont {Devoret}}, \bibinfo {author} {\bibfnamefont {S.~M.}\
			\bibnamefont {Girvin}}, \ and\ \bibinfo {author} {\bibfnamefont {R.~J.}\
			\bibnamefont {Schoelkopf}},\ }\href {\doibase 10.1103/PhysRevA.76.042319}
	{\bibfield  {journal} {\bibinfo  {journal} {Phys. Rev. A}\ }\textbf {\bibinfo
			{volume} {76}},\ \bibinfo {pages} {042319} (\bibinfo {year}
		{2007})}\BibitemShut {NoStop}%
	\bibitem [{\citenamefont {Manucharyan}\ \emph {et~al.}(2009)\citenamefont
		{Manucharyan}, \citenamefont {Koch}, \citenamefont {Glazman},\ and\
		\citenamefont {Devoret}}]{fluxonium}%
	\BibitemOpen
	\bibfield  {author} {\bibinfo {author} {\bibfnamefont {V.~E.}\ \bibnamefont
			{Manucharyan}}, \bibinfo {author} {\bibfnamefont {J.}~\bibnamefont {Koch}},
		\bibinfo {author} {\bibfnamefont {L.~I.}\ \bibnamefont {Glazman}}, \ and\
		\bibinfo {author} {\bibfnamefont {M.~H.}\ \bibnamefont {Devoret}},\ }\href
	{\doibase 10.1126/science.1175552} {\bibfield  {journal} {\bibinfo  {journal}
			{Science}\ }\textbf {\bibinfo {volume} {326}},\ \bibinfo {pages} {113}
		(\bibinfo {year} {2009})}\BibitemShut {NoStop}%
	\bibitem [{\citenamefont {Stern}\ \emph {et~al.}(2014)\citenamefont {Stern},
		\citenamefont {Catelani}, \citenamefont {Kubo}, \citenamefont {Grezes},
		\citenamefont {Bienfait}, \citenamefont {Vion}, \citenamefont {Esteve},\ and\
		\citenamefont {Bertet}}]{Flux-qubit-coherence}%
	\BibitemOpen
	\bibfield  {author} {\bibinfo {author} {\bibfnamefont {M.}~\bibnamefont
			{Stern}}, \bibinfo {author} {\bibfnamefont {G.}~\bibnamefont {Catelani}},
		\bibinfo {author} {\bibfnamefont {Y.}~\bibnamefont {Kubo}}, \bibinfo {author}
		{\bibfnamefont {C.}~\bibnamefont {Grezes}}, \bibinfo {author} {\bibfnamefont
			{A.}~\bibnamefont {Bienfait}}, \bibinfo {author} {\bibfnamefont
			{D.}~\bibnamefont {Vion}}, \bibinfo {author} {\bibfnamefont {D.}~\bibnamefont
			{Esteve}}, \ and\ \bibinfo {author} {\bibfnamefont {P.}~\bibnamefont
			{Bertet}},\ }\href {\doibase 10.1103/PhysRevLett.113.123601} {\bibfield
		{journal} {\bibinfo  {journal} {Phys. Rev. Lett.}\ }\textbf {\bibinfo
			{volume} {113}},\ \bibinfo {pages} {123601} (\bibinfo {year}
		{2014})}\BibitemShut {NoStop}%
	\bibitem [{\citenamefont {Yan}\ \emph {et~al.}(2016)\citenamefont {Yan},
		\citenamefont {Gustavsson}, \citenamefont {Kamal}, \citenamefont {Birenbaum},
		\citenamefont {Sears}, \citenamefont {Hover}, \citenamefont {Gudmundsen},
		\citenamefont {Rosenberg}, \citenamefont {Samach}, \citenamefont {Weber}
		\emph {et~al.}}]{fluxqubit_new}%
	\BibitemOpen
	\bibfield  {author} {\bibinfo {author} {\bibfnamefont {F.}~\bibnamefont
			{Yan}}, \bibinfo {author} {\bibfnamefont {S.}~\bibnamefont {Gustavsson}},
		\bibinfo {author} {\bibfnamefont {A.}~\bibnamefont {Kamal}}, \bibinfo
		{author} {\bibfnamefont {J.}~\bibnamefont {Birenbaum}}, \bibinfo {author}
		{\bibfnamefont {A.~P.}\ \bibnamefont {Sears}}, \bibinfo {author}
		{\bibfnamefont {D.}~\bibnamefont {Hover}}, \bibinfo {author} {\bibfnamefont
			{T.~J.}\ \bibnamefont {Gudmundsen}}, \bibinfo {author} {\bibfnamefont
			{D.}~\bibnamefont {Rosenberg}}, \bibinfo {author} {\bibfnamefont
			{G.}~\bibnamefont {Samach}}, \bibinfo {author} {\bibfnamefont
			{S.}~\bibnamefont {Weber}},  \emph {et~al.},\ }\href {\doibase
		10.1038/ncomms12964} {\bibfield  {journal} {\bibinfo  {journal} {Nature
				communications}\ }\textbf {\bibinfo {volume} {7}},\ \bibinfo {pages} {12964}
		(\bibinfo {year} {2016})}\BibitemShut {NoStop}%
	\bibitem [{\citenamefont {Kelly}\ \emph {et~al.}(2015)\citenamefont {Kelly},
		\citenamefont {Barends}, \citenamefont {Fowler}, \citenamefont {Megrant},
		\citenamefont {Jeffrey}, \citenamefont {White}, \citenamefont {Sank},
		\citenamefont {Mutus}, \citenamefont {Campbell}, \citenamefont {Chen},
		\citenamefont {Chen}, \citenamefont {Chiaro}, \citenamefont {Dunsworth},
		\citenamefont {Hoi}, \citenamefont {Neill}, \citenamefont {O'Malley},
		\citenamefont {Quintana}, \citenamefont {Roushan}, \citenamefont
		{Vainsencher}, \citenamefont {Wenner}, \citenamefont {Cleland},\ and\
		\citenamefont {Martinis}}]{Martinis_9xmon_nature}%
	\BibitemOpen
	\bibfield  {author} {\bibinfo {author} {\bibfnamefont {J.}~\bibnamefont
			{Kelly}}, \bibinfo {author} {\bibfnamefont {R.}~\bibnamefont {Barends}},
		\bibinfo {author} {\bibfnamefont {A.}~\bibnamefont {Fowler}}, \bibinfo
		{author} {\bibfnamefont {A.}~\bibnamefont {Megrant}}, \bibinfo {author}
		{\bibfnamefont {E.}~\bibnamefont {Jeffrey}}, \bibinfo {author} {\bibfnamefont
			{T.}~\bibnamefont {White}}, \bibinfo {author} {\bibfnamefont
			{D.}~\bibnamefont {Sank}}, \bibinfo {author} {\bibfnamefont {J.}~\bibnamefont
			{Mutus}}, \bibinfo {author} {\bibfnamefont {B.}~\bibnamefont {Campbell}},
		\bibinfo {author} {\bibfnamefont {Y.}~\bibnamefont {Chen}}, \bibinfo {author}
		{\bibfnamefont {Z.}~\bibnamefont {Chen}}, \bibinfo {author} {\bibfnamefont
			{B.}~\bibnamefont {Chiaro}}, \bibinfo {author} {\bibfnamefont
			{A.}~\bibnamefont {Dunsworth}}, \bibinfo {author} {\bibfnamefont {I.-C.}\
			\bibnamefont {Hoi}}, \bibinfo {author} {\bibfnamefont {C.}~\bibnamefont
			{Neill}}, \bibinfo {author} {\bibfnamefont {P.}~\bibnamefont {O'Malley}},
		\bibinfo {author} {\bibfnamefont {C.}~\bibnamefont {Quintana}}, \bibinfo
		{author} {\bibfnamefont {P.}~\bibnamefont {Roushan}}, \bibinfo {author}
		{\bibfnamefont {A.}~\bibnamefont {Vainsencher}}, \bibinfo {author}
		{\bibfnamefont {J.}~\bibnamefont {Wenner}}, \bibinfo {author} {\bibfnamefont
			{A.}~\bibnamefont {Cleland}}, \ and\ \bibinfo {author} {\bibfnamefont
			{J.~M.}\ \bibnamefont {Martinis}},\ }\href {\doibase 10.1038/nature14270}
	{\bibfield  {journal} {\bibinfo  {journal} {Nature}\ }\textbf {\bibinfo
			{volume} {519}},\ \bibinfo {pages} {66} (\bibinfo {year} {2015})}\BibitemShut
	{NoStop}%
	\bibitem [{\citenamefont {C{\'o}rcoles}\ \emph {et~al.}(2015)\citenamefont
		{C{\'o}rcoles}, \citenamefont {Magesan}, \citenamefont {Srinivasan},
		\citenamefont {Cross}, \citenamefont {Steffen}, \citenamefont {Gambetta},\
		and\ \citenamefont {Chow}}]{IBM-errordet-4qubit}%
	\BibitemOpen
	\bibfield  {author} {\bibinfo {author} {\bibfnamefont {A.~D.}\ \bibnamefont
			{C{\'o}rcoles}}, \bibinfo {author} {\bibfnamefont {E.}~\bibnamefont
			{Magesan}}, \bibinfo {author} {\bibfnamefont {S.~J.}\ \bibnamefont
			{Srinivasan}}, \bibinfo {author} {\bibfnamefont {A.~W.}\ \bibnamefont
			{Cross}}, \bibinfo {author} {\bibfnamefont {M.}~\bibnamefont {Steffen}},
		\bibinfo {author} {\bibfnamefont {J.~M.}\ \bibnamefont {Gambetta}}, \ and\
		\bibinfo {author} {\bibfnamefont {J.~M.}\ \bibnamefont {Chow}},\ }\href
	{\doibase 10.1038/ncomms7979} {\bibfield  {journal} {\bibinfo  {journal}
			{Nature communications}\ }\textbf {\bibinfo {volume} {6:6979}} (\bibinfo
		{year} {2015}),\ 10.1038/ncomms7979}\BibitemShut {NoStop}%
	\bibitem [{\citenamefont {Rist{\`e}}\ \emph {et~al.}(2015)\citenamefont
		{Rist{\`e}}, \citenamefont {Poletto}, \citenamefont {Huang}, \citenamefont
		{Bruno}, \citenamefont {Vesterinen}, \citenamefont {Saira},\ and\
		\citenamefont {DiCarlo}}]{riste-bit-flip}%
	\BibitemOpen
	\bibfield  {author} {\bibinfo {author} {\bibfnamefont {D.}~\bibnamefont
			{Rist{\`e}}}, \bibinfo {author} {\bibfnamefont {S.}~\bibnamefont {Poletto}},
		\bibinfo {author} {\bibfnamefont {M.-Z.}\ \bibnamefont {Huang}}, \bibinfo
		{author} {\bibfnamefont {A.}~\bibnamefont {Bruno}}, \bibinfo {author}
		{\bibfnamefont {V.}~\bibnamefont {Vesterinen}}, \bibinfo {author}
		{\bibfnamefont {O.-P.}\ \bibnamefont {Saira}}, \ and\ \bibinfo {author}
		{\bibfnamefont {L.}~\bibnamefont {DiCarlo}},\ }\href {\doibase
		10.1038/ncomms7983} {\bibfield  {journal} {\bibinfo  {journal} {Nature
				communications}\ }\textbf {\bibinfo {volume} {6:6983}} (\bibinfo {year}
		{2015}),\ 10.1038/ncomms7983}\BibitemShut {NoStop}%
	\bibitem [{\citenamefont {Blumoff}\ \emph {et~al.}(2016)\citenamefont
		{Blumoff}, \citenamefont {Chou}, \citenamefont {Shen}, \citenamefont
		{Reagor}, \citenamefont {Axline}, \citenamefont {Brierley}, \citenamefont
		{Silveri}, \citenamefont {Wang}, \citenamefont {Vlastakis}, \citenamefont
		{Nigg}, \citenamefont {Frunzio}, \citenamefont {Devoret}, \citenamefont
		{Jiang}, \citenamefont {Girvin},\ and\ \citenamefont
		{Schoelkopf}}]{multi-qubit-3d-yale}%
	\BibitemOpen
	\bibfield  {author} {\bibinfo {author} {\bibfnamefont {J.~Z.}\ \bibnamefont
			{Blumoff}}, \bibinfo {author} {\bibfnamefont {K.}~\bibnamefont {Chou}},
		\bibinfo {author} {\bibfnamefont {C.}~\bibnamefont {Shen}}, \bibinfo {author}
		{\bibfnamefont {M.}~\bibnamefont {Reagor}}, \bibinfo {author} {\bibfnamefont
			{C.}~\bibnamefont {Axline}}, \bibinfo {author} {\bibfnamefont {R.~T.}\
			\bibnamefont {Brierley}}, \bibinfo {author} {\bibfnamefont {M.~P.}\
			\bibnamefont {Silveri}}, \bibinfo {author} {\bibfnamefont {C.}~\bibnamefont
			{Wang}}, \bibinfo {author} {\bibfnamefont {B.}~\bibnamefont {Vlastakis}},
		\bibinfo {author} {\bibfnamefont {S.~E.}\ \bibnamefont {Nigg}}, \bibinfo
		{author} {\bibfnamefont {L.}~\bibnamefont {Frunzio}}, \bibinfo {author}
		{\bibfnamefont {M.~H.}\ \bibnamefont {Devoret}}, \bibinfo {author}
		{\bibfnamefont {L.}~\bibnamefont {Jiang}}, \bibinfo {author} {\bibfnamefont
			{S.~M.}\ \bibnamefont {Girvin}}, \ and\ \bibinfo {author} {\bibfnamefont
			{R.~J.}\ \bibnamefont {Schoelkopf}},\ }\href {\doibase
		10.1103/PhysRevX.6.031041} {\bibfield  {journal} {\bibinfo  {journal} {Phys.
				Rev. X}\ }\textbf {\bibinfo {volume} {6}},\ \bibinfo {pages} {031041}
		(\bibinfo {year} {2016})}\BibitemShut {NoStop}%
	\bibitem [{\citenamefont {Barends}\ \emph {et~al.}(2015)\citenamefont
		{Barends}, \citenamefont {Lamata}, \citenamefont {Kelly}, \citenamefont
		{Garc{\'\i}a-{\'A}lvarez}, \citenamefont {Fowler}, \citenamefont {Megrant},
		\citenamefont {Jeffrey}, \citenamefont {White}, \citenamefont {Sank},
		\citenamefont {Mutus}, \citenamefont {Campbell}, \citenamefont {Chen},
		\citenamefont {Chen}, \citenamefont {Chiaro}, \citenamefont {Dunsworth},
		\citenamefont {Hoi}, \citenamefont {Neill}, \citenamefont {O’Malley},
		\citenamefont {Quintana}, \citenamefont {Roushan}, \citenamefont
		{Vainsencher}, \citenamefont {Wenner}, \citenamefont {Solano},\ and\
		\citenamefont {Martinis}}]{Qsim_xmon}%
	\BibitemOpen
	\bibfield  {author} {\bibinfo {author} {\bibfnamefont {R.}~\bibnamefont
			{Barends}}, \bibinfo {author} {\bibfnamefont {L.}~\bibnamefont {Lamata}},
		\bibinfo {author} {\bibfnamefont {J.}~\bibnamefont {Kelly}}, \bibinfo
		{author} {\bibfnamefont {L.}~\bibnamefont {Garc{\'\i}a-{\'A}lvarez}},
		\bibinfo {author} {\bibfnamefont {A.}~\bibnamefont {Fowler}}, \bibinfo
		{author} {\bibfnamefont {A.}~\bibnamefont {Megrant}}, \bibinfo {author}
		{\bibfnamefont {E.}~\bibnamefont {Jeffrey}}, \bibinfo {author} {\bibfnamefont
			{T.}~\bibnamefont {White}}, \bibinfo {author} {\bibfnamefont
			{D.}~\bibnamefont {Sank}}, \bibinfo {author} {\bibfnamefont {J.}~\bibnamefont
			{Mutus}}, \bibinfo {author} {\bibfnamefont {B.}~\bibnamefont {Campbell}},
		\bibinfo {author} {\bibfnamefont {Y.}~\bibnamefont {Chen}}, \bibinfo {author}
		{\bibfnamefont {Z.}~\bibnamefont {Chen}}, \bibinfo {author} {\bibfnamefont
			{B.}~\bibnamefont {Chiaro}}, \bibinfo {author} {\bibfnamefont
			{A.}~\bibnamefont {Dunsworth}}, \bibinfo {author} {\bibfnamefont {I.-C.}\
			\bibnamefont {Hoi}}, \bibinfo {author} {\bibfnamefont {C.}~\bibnamefont
			{Neill}}, \bibinfo {author} {\bibfnamefont {P.~J.~J.}\ \bibnamefont
			{O’Malley}}, \bibinfo {author} {\bibfnamefont {C.}~\bibnamefont
			{Quintana}}, \bibinfo {author} {\bibfnamefont {P.}~\bibnamefont {Roushan}},
		\bibinfo {author} {\bibfnamefont {A.}~\bibnamefont {Vainsencher}}, \bibinfo
		{author} {\bibfnamefont {J.}~\bibnamefont {Wenner}}, \bibinfo {author}
		{\bibfnamefont {E.}~\bibnamefont {Solano}}, \ and\ \bibinfo {author}
		{\bibfnamefont {J.~M.}\ \bibnamefont {Martinis}},\ }\href {\doibase
		10.1038/ncomms8654} {\bibfield  {journal} {\bibinfo  {journal} {Nature
				communications}\ }\textbf {\bibinfo {volume} {6:7654}} (\bibinfo {year}
		{2015}),\ 10.1038/ncomms8654}\BibitemShut {NoStop}%
	\bibitem [{\citenamefont {Salath\'e}\ \emph {et~al.}(2015)\citenamefont
		{Salath\'e}, \citenamefont {Mondal}, \citenamefont {Oppliger}, \citenamefont
		{Heinsoo}, \citenamefont {Kurpiers}, \citenamefont
		{Poto\ifmmode~\check{c}\else \v{c}\fi{}nik}, \citenamefont {Mezzacapo},
		\citenamefont {Las~Heras}, \citenamefont {Lamata}, \citenamefont {Solano},
		\citenamefont {Filipp},\ and\ \citenamefont {Wallraff}}]{wallraff-dig-qsim}%
	\BibitemOpen
	\bibfield  {author} {\bibinfo {author} {\bibfnamefont {Y.}~\bibnamefont
			{Salath\'e}}, \bibinfo {author} {\bibfnamefont {M.}~\bibnamefont {Mondal}},
		\bibinfo {author} {\bibfnamefont {M.}~\bibnamefont {Oppliger}}, \bibinfo
		{author} {\bibfnamefont {J.}~\bibnamefont {Heinsoo}}, \bibinfo {author}
		{\bibfnamefont {P.}~\bibnamefont {Kurpiers}}, \bibinfo {author}
		{\bibfnamefont {A.}~\bibnamefont {Poto\ifmmode~\check{c}\else
				\v{c}\fi{}nik}}, \bibinfo {author} {\bibfnamefont {A.}~\bibnamefont
			{Mezzacapo}}, \bibinfo {author} {\bibfnamefont {U.}~\bibnamefont
			{Las~Heras}}, \bibinfo {author} {\bibfnamefont {L.}~\bibnamefont {Lamata}},
		\bibinfo {author} {\bibfnamefont {E.}~\bibnamefont {Solano}}, \bibinfo
		{author} {\bibfnamefont {S.}~\bibnamefont {Filipp}}, \ and\ \bibinfo {author}
		{\bibfnamefont {A.}~\bibnamefont {Wallraff}},\ }\href {\doibase
		10.1103/PhysRevX.5.021027} {\bibfield  {journal} {\bibinfo  {journal} {Phys.
				Rev. X}\ }\textbf {\bibinfo {volume} {5}},\ \bibinfo {pages} {021027}
		(\bibinfo {year} {2015})}\BibitemShut {NoStop}%
	\bibitem [{\citenamefont {Paik}\ \emph {et~al.}(2016)\citenamefont {Paik},
		\citenamefont {Mezzacapo}, \citenamefont {Sandberg}, \citenamefont {McClure},
		\citenamefont {Abdo}, \citenamefont {C\'orcoles}, \citenamefont {Dial},
		\citenamefont {Bogorin}, \citenamefont {Plourde}, \citenamefont {Steffen},
		\citenamefont {Cross}, \citenamefont {Gambetta},\ and\ \citenamefont
		{Chow}}]{RIP-gate}%
	\BibitemOpen
	\bibfield  {author} {\bibinfo {author} {\bibfnamefont {H.}~\bibnamefont
			{Paik}}, \bibinfo {author} {\bibfnamefont {A.}~\bibnamefont {Mezzacapo}},
		\bibinfo {author} {\bibfnamefont {M.}~\bibnamefont {Sandberg}}, \bibinfo
		{author} {\bibfnamefont {D.~T.}\ \bibnamefont {McClure}}, \bibinfo {author}
		{\bibfnamefont {B.}~\bibnamefont {Abdo}}, \bibinfo {author} {\bibfnamefont
			{A.~D.}\ \bibnamefont {C\'orcoles}}, \bibinfo {author} {\bibfnamefont
			{O.}~\bibnamefont {Dial}}, \bibinfo {author} {\bibfnamefont {D.~F.}\
			\bibnamefont {Bogorin}}, \bibinfo {author} {\bibfnamefont {B.~L.~T.}\
			\bibnamefont {Plourde}}, \bibinfo {author} {\bibfnamefont {M.}~\bibnamefont
			{Steffen}}, \bibinfo {author} {\bibfnamefont {A.~W.}\ \bibnamefont {Cross}},
		\bibinfo {author} {\bibfnamefont {J.~M.}\ \bibnamefont {Gambetta}}, \ and\
		\bibinfo {author} {\bibfnamefont {J.~M.}\ \bibnamefont {Chow}},\ }\href
	{\doibase 10.1103/PhysRevLett.117.250502} {\bibfield  {journal} {\bibinfo
			{journal} {Phys. Rev. Lett.}\ }\textbf {\bibinfo {volume} {117}},\ \bibinfo
		{pages} {250502} (\bibinfo {year} {2016})}\BibitemShut {NoStop}%
	\bibitem [{\citenamefont {Roy}\ \emph {et~al.}(2017)\citenamefont {Roy},
		\citenamefont {Kundu}, \citenamefont {Chand}, \citenamefont {Hazra},
		\citenamefont {Nehra}, \citenamefont {Cosmic}, \citenamefont {Ranadive},
		\citenamefont {Patankar}, \citenamefont {Damle},\ and\ \citenamefont
		{Vijay}}]{trimon}%
	\BibitemOpen
	\bibfield  {author} {\bibinfo {author} {\bibfnamefont {T.}~\bibnamefont
			{Roy}}, \bibinfo {author} {\bibfnamefont {S.}~\bibnamefont {Kundu}}, \bibinfo
		{author} {\bibfnamefont {M.}~\bibnamefont {Chand}}, \bibinfo {author}
		{\bibfnamefont {S.}~\bibnamefont {Hazra}}, \bibinfo {author} {\bibfnamefont
			{N.}~\bibnamefont {Nehra}}, \bibinfo {author} {\bibfnamefont
			{R.}~\bibnamefont {Cosmic}}, \bibinfo {author} {\bibfnamefont
			{A.}~\bibnamefont {Ranadive}}, \bibinfo {author} {\bibfnamefont {M.~P.}\
			\bibnamefont {Patankar}}, \bibinfo {author} {\bibfnamefont {K.}~\bibnamefont
			{Damle}}, \ and\ \bibinfo {author} {\bibfnamefont {R.}~\bibnamefont
			{Vijay}},\ }\href {\doibase 10.1103/PhysRevApplied.7.054025} {\bibfield
		{journal} {\bibinfo  {journal} {Phys. Rev. Applied}\ }\textbf {\bibinfo
			{volume} {7}},\ \bibinfo {pages} {054025} (\bibinfo {year}
		{2017})}\BibitemShut {NoStop}%
	\bibitem [{\citenamefont {Kou}\ \emph {et~al.}(2017)\citenamefont {Kou},
		\citenamefont {Smith}, \citenamefont {Vool}, \citenamefont {Brierley},
		\citenamefont {Meier}, \citenamefont {Frunzio}, \citenamefont {Girvin},
		\citenamefont {Glazman},\ and\ \citenamefont {Devoret}}]{fluxonium_molecule}%
	\BibitemOpen
	\bibfield  {author} {\bibinfo {author} {\bibfnamefont {A.}~\bibnamefont
			{Kou}}, \bibinfo {author} {\bibfnamefont {W.~C.}\ \bibnamefont {Smith}},
		\bibinfo {author} {\bibfnamefont {U.}~\bibnamefont {Vool}}, \bibinfo {author}
		{\bibfnamefont {R.~T.}\ \bibnamefont {Brierley}}, \bibinfo {author}
		{\bibfnamefont {H.}~\bibnamefont {Meier}}, \bibinfo {author} {\bibfnamefont
			{L.}~\bibnamefont {Frunzio}}, \bibinfo {author} {\bibfnamefont {S.~M.}\
			\bibnamefont {Girvin}}, \bibinfo {author} {\bibfnamefont {L.~I.}\
			\bibnamefont {Glazman}}, \ and\ \bibinfo {author} {\bibfnamefont {M.~H.}\
			\bibnamefont {Devoret}},\ }\href {\doibase 10.1103/PhysRevX.7.031037}
	{\bibfield  {journal} {\bibinfo  {journal} {Phys. Rev. X}\ }\textbf {\bibinfo
			{volume} {7}},\ \bibinfo {pages} {031037} (\bibinfo {year}
		{2017})}\BibitemShut {NoStop}%
	\bibitem [{\citenamefont {Hoffman}\ \emph {et~al.}(2011)\citenamefont
		{Hoffman}, \citenamefont {Srinivasan}, \citenamefont {Gambetta},\ and\
		\citenamefont {Houck}}]{Houck-TCQ2}%
	\BibitemOpen
	\bibfield  {author} {\bibinfo {author} {\bibfnamefont {A.~J.}\ \bibnamefont
			{Hoffman}}, \bibinfo {author} {\bibfnamefont {S.~J.}\ \bibnamefont
			{Srinivasan}}, \bibinfo {author} {\bibfnamefont {J.~M.}\ \bibnamefont
			{Gambetta}}, \ and\ \bibinfo {author} {\bibfnamefont {A.~A.}\ \bibnamefont
			{Houck}},\ }\href {\doibase 10.1103/PhysRevB.84.184515} {\bibfield  {journal}
		{\bibinfo  {journal} {Physical Review B}\ }\textbf {\bibinfo {volume} {84}},\
		\bibinfo {pages} {184515} (\bibinfo {year} {2011})}\BibitemShut {NoStop}%
	\bibitem [{\citenamefont {Dumur}\ \emph {et~al.}(2015)\citenamefont {Dumur},
		\citenamefont {K{\"u}ng}, \citenamefont {Feofanov}, \citenamefont {Weissl},
		\citenamefont {Roch}, \citenamefont {Naud}, \citenamefont {Guichard},\ and\
		\citenamefont {Buisson}}]{Buisson-Vshaped}%
	\BibitemOpen
	\bibfield  {author} {\bibinfo {author} {\bibfnamefont {{\'E}.}~\bibnamefont
			{Dumur}}, \bibinfo {author} {\bibfnamefont {B.}~\bibnamefont {K{\"u}ng}},
		\bibinfo {author} {\bibfnamefont {A.~K.}\ \bibnamefont {Feofanov}}, \bibinfo
		{author} {\bibfnamefont {T.}~\bibnamefont {Weissl}}, \bibinfo {author}
		{\bibfnamefont {N.}~\bibnamefont {Roch}}, \bibinfo {author} {\bibfnamefont
			{C.}~\bibnamefont {Naud}}, \bibinfo {author} {\bibfnamefont {W.}~\bibnamefont
			{Guichard}}, \ and\ \bibinfo {author} {\bibfnamefont {O.}~\bibnamefont
			{Buisson}},\ }\href {\doibase 10.1103/PhysRevB.92.020515} {\bibfield
		{journal} {\bibinfo  {journal} {Physical Review B}\ }\textbf {\bibinfo
			{volume} {92}},\ \bibinfo {pages} {020515} (\bibinfo {year}
		{2015})}\BibitemShut {NoStop}%
	\bibitem [{\citenamefont {Zhang}\ \emph
		{et~al.}(2017{\natexlab{a}})\citenamefont {Zhang}, \citenamefont {Liu},
		\citenamefont {Raftery},\ and\ \citenamefont
		{Houck}}]{houck-TCQ-photonnoise}%
	\BibitemOpen
	\bibfield  {author} {\bibinfo {author} {\bibfnamefont {G.}~\bibnamefont
			{Zhang}}, \bibinfo {author} {\bibfnamefont {Y.}~\bibnamefont {Liu}}, \bibinfo
		{author} {\bibfnamefont {J.~J.}\ \bibnamefont {Raftery}}, \ and\ \bibinfo
		{author} {\bibfnamefont {A.~A.}\ \bibnamefont {Houck}},\ }\href@noop {}
	{\bibfield  {journal} {\bibinfo  {journal} {npj Quantum Information}\
		}\textbf {\bibinfo {volume} {3}},\ \bibinfo {pages} {1} (\bibinfo {year}
		{2017}{\natexlab{a}})}\BibitemShut {NoStop}%
	\bibitem [{\citenamefont {Makhlin}\ \emph {et~al.}(1999)\citenamefont
		{Makhlin}, \citenamefont {Sc{\"o}hn},\ and\ \citenamefont
		{Shnirman}}]{nature-cpb-schnirman}%
	\BibitemOpen
	\bibfield  {author} {\bibinfo {author} {\bibfnamefont {Y.}~\bibnamefont
			{Makhlin}}, \bibinfo {author} {\bibfnamefont {G.}~\bibnamefont {Sc{\"o}hn}},
		\ and\ \bibinfo {author} {\bibfnamefont {A.}~\bibnamefont {Shnirman}},\
	}\href {\doibase 10.1038/18613} {\bibfield  {journal} {\bibinfo  {journal}
			{nature}\ }\textbf {\bibinfo {volume} {398}},\ \bibinfo {pages} {305}
		(\bibinfo {year} {1999})}\BibitemShut {NoStop}%
	\bibitem [{\citenamefont {You}\ \emph {et~al.}(2002)\citenamefont {You},
		\citenamefont {Tsai},\ and\ \citenamefont {Nori}}]{cpb-array-nori}%
	\BibitemOpen
	\bibfield  {author} {\bibinfo {author} {\bibfnamefont {J.~Q.}\ \bibnamefont
			{You}}, \bibinfo {author} {\bibfnamefont {J.~S.}\ \bibnamefont {Tsai}}, \
		and\ \bibinfo {author} {\bibfnamefont {F.}~\bibnamefont {Nori}},\ }\href
	{\doibase 10.1103/PhysRevLett.89.197902} {\bibfield  {journal} {\bibinfo
			{journal} {Phys. Rev. Lett.}\ }\textbf {\bibinfo {volume} {89}},\ \bibinfo
		{pages} {197902} (\bibinfo {year} {2002})}\BibitemShut {NoStop}%
	\bibitem [{\citenamefont {Diniz}\ \emph {et~al.}(2013)\citenamefont {Diniz},
		\citenamefont {Dumur}, \citenamefont {Buisson},\ and\ \citenamefont
		{Auffeves}}]{Auffeves_Buisson-theory}%
	\BibitemOpen
	\bibfield  {author} {\bibinfo {author} {\bibfnamefont {I.}~\bibnamefont
			{Diniz}}, \bibinfo {author} {\bibfnamefont {E.}~\bibnamefont {Dumur}},
		\bibinfo {author} {\bibfnamefont {O.}~\bibnamefont {Buisson}}, \ and\
		\bibinfo {author} {\bibfnamefont {A.}~\bibnamefont {Auffeves}},\ }\href
	{\doibase 10.1103/PhysRevA.87.033837} {\bibfield  {journal} {\bibinfo
			{journal} {Physical Review A}\ }\textbf {\bibinfo {volume} {87}},\ \bibinfo
		{pages} {033837} (\bibinfo {year} {2013})}\BibitemShut {NoStop}%
	\bibitem [{\citenamefont {Nigg}\ \emph {et~al.}(2012)\citenamefont {Nigg},
		\citenamefont {Paik}, \citenamefont {Vlastakis}, \citenamefont {Kirchmair},
		\citenamefont {Shankar}, \citenamefont {Frunzio}, \citenamefont {Devoret},
		\citenamefont {Schoelkopf},\ and\ \citenamefont {Girvin}}]{bbq-yale}%
	\BibitemOpen
	\bibfield  {author} {\bibinfo {author} {\bibfnamefont {S.~E.}\ \bibnamefont
			{Nigg}}, \bibinfo {author} {\bibfnamefont {H.}~\bibnamefont {Paik}}, \bibinfo
		{author} {\bibfnamefont {B.}~\bibnamefont {Vlastakis}}, \bibinfo {author}
		{\bibfnamefont {G.}~\bibnamefont {Kirchmair}}, \bibinfo {author}
		{\bibfnamefont {S.}~\bibnamefont {Shankar}}, \bibinfo {author} {\bibfnamefont
			{L.}~\bibnamefont {Frunzio}}, \bibinfo {author} {\bibfnamefont {M.~H.}\
			\bibnamefont {Devoret}}, \bibinfo {author} {\bibfnamefont {R.~J.}\
			\bibnamefont {Schoelkopf}}, \ and\ \bibinfo {author} {\bibfnamefont {S.~M.}\
			\bibnamefont {Girvin}},\ }\href {\doibase 10.1103/PhysRevLett.108.240502}
	{\bibfield  {journal} {\bibinfo  {journal} {Phys. Rev. Lett.}\ }\textbf
		{\bibinfo {volume} {108}},\ \bibinfo {pages} {240502} (\bibinfo {year}
		{2012})}\BibitemShut {NoStop}%
	\bibitem [{\citenamefont {Bourassa}\ \emph {et~al.}(2012)\citenamefont
		{Bourassa}, \citenamefont {Beaudoin}, \citenamefont {Gambetta},\ and\
		\citenamefont {Blais}}]{quantize-blais}%
	\BibitemOpen
	\bibfield  {author} {\bibinfo {author} {\bibfnamefont {J.}~\bibnamefont
			{Bourassa}}, \bibinfo {author} {\bibfnamefont {F.}~\bibnamefont {Beaudoin}},
		\bibinfo {author} {\bibfnamefont {J.~M.}\ \bibnamefont {Gambetta}}, \ and\
		\bibinfo {author} {\bibfnamefont {A.}~\bibnamefont {Blais}},\ }\href
	{\doibase 10.1103/PhysRevA.86.013814} {\bibfield  {journal} {\bibinfo
			{journal} {Phys. Rev. A}\ }\textbf {\bibinfo {volume} {86}},\ \bibinfo
		{pages} {013814} (\bibinfo {year} {2012})}\BibitemShut {NoStop}%
	\bibitem [{\citenamefont {Solgun}\ \emph {et~al.}(2014)\citenamefont {Solgun},
		\citenamefont {Abraham},\ and\ \citenamefont {DiVincenzo}}]{bbq-divincenzo}%
	\BibitemOpen
	\bibfield  {author} {\bibinfo {author} {\bibfnamefont {F.}~\bibnamefont
			{Solgun}}, \bibinfo {author} {\bibfnamefont {D.~W.}\ \bibnamefont {Abraham}},
		\ and\ \bibinfo {author} {\bibfnamefont {D.~P.}\ \bibnamefont {DiVincenzo}},\
	}\href {\doibase 10.1103/PhysRevB.90.134504} {\bibfield  {journal} {\bibinfo
			{journal} {Phys. Rev. B}\ }\textbf {\bibinfo {volume} {90}},\ \bibinfo
		{pages} {134504} (\bibinfo {year} {2014})}\BibitemShut {NoStop}%
	\bibitem [{\citenamefont {Wei\ss{}l}\ \emph {et~al.}(2015)\citenamefont
		{Wei\ss{}l}, \citenamefont {K\"ung}, \citenamefont {Dumur}, \citenamefont
		{Feofanov}, \citenamefont {Matei}, \citenamefont {Naud}, \citenamefont
		{Buisson}, \citenamefont {Hekking},\ and\ \citenamefont
		{Guichard}}]{Kerr-coeff}%
	\BibitemOpen
	\bibfield  {author} {\bibinfo {author} {\bibfnamefont {T.}~\bibnamefont
			{Wei\ss{}l}}, \bibinfo {author} {\bibfnamefont {B.}~\bibnamefont {K\"ung}},
		\bibinfo {author} {\bibfnamefont {E.}~\bibnamefont {Dumur}}, \bibinfo
		{author} {\bibfnamefont {A.~K.}\ \bibnamefont {Feofanov}}, \bibinfo {author}
		{\bibfnamefont {I.}~\bibnamefont {Matei}}, \bibinfo {author} {\bibfnamefont
			{C.}~\bibnamefont {Naud}}, \bibinfo {author} {\bibfnamefont {O.}~\bibnamefont
			{Buisson}}, \bibinfo {author} {\bibfnamefont {F.~W.~J.}\ \bibnamefont
			{Hekking}}, \ and\ \bibinfo {author} {\bibfnamefont {W.}~\bibnamefont
			{Guichard}},\ }\href {\doibase 10.1103/PhysRevB.92.104508} {\bibfield
		{journal} {\bibinfo  {journal} {Phys. Rev. B}\ }\textbf {\bibinfo {volume}
			{92}},\ \bibinfo {pages} {104508} (\bibinfo {year} {2015})}\BibitemShut
	{NoStop}%
	\bibitem [{\citenamefont {Bergeal}\ \emph {et~al.}(2010)\citenamefont
		{Bergeal}, \citenamefont {Vijay}, \citenamefont {Manucharyan}, \citenamefont
		{Siddiqi}, \citenamefont {Schoelkopf}, \citenamefont {Girvin},\ and\
		\citenamefont {Devoret}}]{JRM}%
	\BibitemOpen
	\bibfield  {author} {\bibinfo {author} {\bibfnamefont {N.}~\bibnamefont
			{Bergeal}}, \bibinfo {author} {\bibfnamefont {R.}~\bibnamefont {Vijay}},
		\bibinfo {author} {\bibfnamefont {V.~E.}\ \bibnamefont {Manucharyan}},
		\bibinfo {author} {\bibfnamefont {I.}~\bibnamefont {Siddiqi}}, \bibinfo
		{author} {\bibfnamefont {R.~J.}\ \bibnamefont {Schoelkopf}}, \bibinfo
		{author} {\bibfnamefont {S.~M.}\ \bibnamefont {Girvin}}, \ and\ \bibinfo
		{author} {\bibfnamefont {M.~H.}\ \bibnamefont {Devoret}},\ }\href {\doibase
		10.1038/nphys1516} {\bibfield  {journal} {\bibinfo  {journal} {Nature
				Physics}\ }\textbf {\bibinfo {volume} {6}},\ \bibinfo {pages} {296} (\bibinfo
		{year} {2010})}\BibitemShut {NoStop}%
	\bibitem [{\citenamefont {Gambetta}\ \emph {et~al.}(2011)\citenamefont
		{Gambetta}, \citenamefont {Houck},\ and\ \citenamefont {Blais}}]{Blais_TCQ}%
	\BibitemOpen
	\bibfield  {author} {\bibinfo {author} {\bibfnamefont {J.~M.}\ \bibnamefont
			{Gambetta}}, \bibinfo {author} {\bibfnamefont {A.~A.}\ \bibnamefont {Houck}},
		\ and\ \bibinfo {author} {\bibfnamefont {A.}~\bibnamefont {Blais}},\ }\href
	{\doibase 10.1103/PhysRevLett.106.030502} {\bibfield  {journal} {\bibinfo
			{journal} {Physical review letters}\ }\textbf {\bibinfo {volume} {106}},\
		\bibinfo {pages} {030502} (\bibinfo {year} {2011})}\BibitemShut {NoStop}%
	\bibitem [{\citenamefont {Leib}\ \emph {et~al.}(2016)\citenamefont {Leib},
		\citenamefont {Zoller},\ and\ \citenamefont
		{Lechner}}]{Transmon-annealer-Zoller}%
	\BibitemOpen
	\bibfield  {author} {\bibinfo {author} {\bibfnamefont {M.}~\bibnamefont
			{Leib}}, \bibinfo {author} {\bibfnamefont {P.}~\bibnamefont {Zoller}}, \ and\
		\bibinfo {author} {\bibfnamefont {W.}~\bibnamefont {Lechner}},\ }\href
	{http://stacks.iop.org/2058-9565/1/i=1/a=015008} {\bibfield  {journal}
		{\bibinfo  {journal} {Quantum Science and Technology}\ }\textbf {\bibinfo
			{volume} {1}},\ \bibinfo {pages} {015008} (\bibinfo {year}
		{2016})}\BibitemShut {NoStop}%
	\bibitem [{\citenamefont {Zhang}\ \emph
		{et~al.}(2017{\natexlab{b}})\citenamefont {Zhang}, \citenamefont {Liu},
		\citenamefont {Raftery},\ and\ \citenamefont {Houck}}]{tcq-houck}%
	\BibitemOpen
	\bibfield  {author} {\bibinfo {author} {\bibfnamefont {G.}~\bibnamefont
			{Zhang}}, \bibinfo {author} {\bibfnamefont {Y.}~\bibnamefont {Liu}}, \bibinfo
		{author} {\bibfnamefont {J.~J.}\ \bibnamefont {Raftery}}, \ and\ \bibinfo
		{author} {\bibfnamefont {A.~A.}\ \bibnamefont {Houck}},\ }\href {\doibase
		doi:10.1038/s41534-016-0002-2} {\bibfield  {journal} {\bibinfo  {journal}
			{npj Quantum Information}\ }\textbf {\bibinfo {volume} {3}} (\bibinfo {year}
		{2017}{\natexlab{b}}),\ doi:10.1038/s41534-016-0002-2}\BibitemShut {NoStop}%
	\bibitem [{\citenamefont {Flurin}(2014)}]{flurin_thesis}%
	\BibitemOpen
	\bibfield  {author} {\bibinfo {author} {\bibfnamefont {E.}~\bibnamefont
			{Flurin}},\ }\emph {\bibinfo {title} {{The Josephson Mixer, a Swiss army
				knife for microwave quantum optics}}},\ \href
	{https://tel.archives-ouvertes.fr/tel-01241123} {\bibinfo {type} {Theses}},\
	\bibinfo  {school} {{Ecole Normale Sup{\'e}rieure, Paris}} (\bibinfo {year}
	{2014})\BibitemShut {NoStop}%
	\bibitem [{\citenamefont {Pop}\ \emph {et~al.}(2010)\citenamefont {Pop},
		\citenamefont {Protopopov}, \citenamefont {Lecocq}, \citenamefont {Peng},
		\citenamefont {Pannetier}, \citenamefont {Buisson},\ and\ \citenamefont
		{Guichard}}]{phase-slip_chain}%
	\BibitemOpen
	\bibfield  {author} {\bibinfo {author} {\bibfnamefont {I.~M.}\ \bibnamefont
			{Pop}}, \bibinfo {author} {\bibfnamefont {I.}~\bibnamefont {Protopopov}},
		\bibinfo {author} {\bibfnamefont {F.}~\bibnamefont {Lecocq}}, \bibinfo
		{author} {\bibfnamefont {Z.}~\bibnamefont {Peng}}, \bibinfo {author}
		{\bibfnamefont {B.}~\bibnamefont {Pannetier}}, \bibinfo {author}
		{\bibfnamefont {O.}~\bibnamefont {Buisson}}, \ and\ \bibinfo {author}
		{\bibfnamefont {W.}~\bibnamefont {Guichard}},\ }\href {\doibase
		10.1038/nphys1697} {\bibfield  {journal} {\bibinfo  {journal} {Nat. Phys.}\
		}\textbf {\bibinfo {volume} {6}} (\bibinfo {year} {2010}),\
		10.1038/nphys1697}\BibitemShut {NoStop}%
	\bibitem [{\citenamefont {Pop}\ \emph {et~al.}(2008)\citenamefont {Pop},
		\citenamefont {Hasselbach}, \citenamefont {Buisson}, \citenamefont
		{Guichard}, \citenamefont {Pannetier},\ and\ \citenamefont
		{Protopopov}}]{phase-slip_rhombi}%
	\BibitemOpen
	\bibfield  {author} {\bibinfo {author} {\bibfnamefont {I.~M.}\ \bibnamefont
			{Pop}}, \bibinfo {author} {\bibfnamefont {K.}~\bibnamefont {Hasselbach}},
		\bibinfo {author} {\bibfnamefont {O.}~\bibnamefont {Buisson}}, \bibinfo
		{author} {\bibfnamefont {W.}~\bibnamefont {Guichard}}, \bibinfo {author}
		{\bibfnamefont {B.}~\bibnamefont {Pannetier}}, \ and\ \bibinfo {author}
		{\bibfnamefont {I.}~\bibnamefont {Protopopov}},\ }\href {\doibase
		10.1103/PhysRevB.78.104504} {\bibfield  {journal} {\bibinfo  {journal} {Phys.
				Rev. B}\ }\textbf {\bibinfo {volume} {78}},\ \bibinfo {pages} {104504}
		(\bibinfo {year} {2008})}\BibitemShut {NoStop}%
	\bibitem [{\citenamefont {Fredkin}\ and\ \citenamefont
		{Toffoli}(1982)}]{Fredkin1982}%
	\BibitemOpen
	\bibfield  {author} {\bibinfo {author} {\bibfnamefont {E.}~\bibnamefont
			{Fredkin}}\ and\ \bibinfo {author} {\bibfnamefont {T.}~\bibnamefont
			{Toffoli}},\ }\href {\doibase 10.1007/BF01857727} {\bibfield  {journal}
		{\bibinfo  {journal} {International Journal of Theoretical Physics}\ }\textbf
		{\bibinfo {volume} {21}},\ \bibinfo {pages} {219} (\bibinfo {year}
		{1982})}\BibitemShut {NoStop}%
	\bibitem [{\citenamefont {Vandersypen}\ and\ \citenamefont
		{Chuang}(2005)}]{NMR_technique_Chuang}%
	\BibitemOpen
	\bibfield  {author} {\bibinfo {author} {\bibfnamefont {L.~M.~K.}\
			\bibnamefont {Vandersypen}}\ and\ \bibinfo {author} {\bibfnamefont {I.~L.}\
			\bibnamefont {Chuang}},\ }\href {\doibase 10.1103/RevModPhys.76.1037}
	{\bibfield  {journal} {\bibinfo  {journal} {Rev. Mod. Phys.}\ }\textbf
		{\bibinfo {volume} {76}},\ \bibinfo {pages} {1037} (\bibinfo {year}
		{2005})}\BibitemShut {NoStop}%
	\bibitem [{\citenamefont {McKay}\ \emph {et~al.}(2017)\citenamefont {McKay},
		\citenamefont {Wood}, \citenamefont {Sheldon}, \citenamefont {Chow},\ and\
		\citenamefont {Gambetta}}]{ibm_zgate_free}%
	\BibitemOpen
	\bibfield  {author} {\bibinfo {author} {\bibfnamefont {D.~C.}\ \bibnamefont
			{McKay}}, \bibinfo {author} {\bibfnamefont {C.~J.}\ \bibnamefont {Wood}},
		\bibinfo {author} {\bibfnamefont {S.}~\bibnamefont {Sheldon}}, \bibinfo
		{author} {\bibfnamefont {J.~M.}\ \bibnamefont {Chow}}, \ and\ \bibinfo
		{author} {\bibfnamefont {J.~M.}\ \bibnamefont {Gambetta}},\ }\href {\doibase
		10.1103/PhysRevA.96.022330} {\bibfield  {journal} {\bibinfo  {journal} {Phys.
				Rev. A}\ }\textbf {\bibinfo {volume} {96}},\ \bibinfo {pages} {022330}
		(\bibinfo {year} {2017})}\BibitemShut {NoStop}%
	\bibitem [{\citenamefont {Riste}\ \emph {et~al.}(2013)\citenamefont {Riste},
		\citenamefont {Dukalski}, \citenamefont {Watson}, \citenamefont {De~Lange},
		\citenamefont {Tiggelman}, \citenamefont {Blanter}, \citenamefont {Lehnert},
		\citenamefont {Schouten},\ and\ \citenamefont {DiCarlo}}]{JPA-parity}%
	\BibitemOpen
	\bibfield  {author} {\bibinfo {author} {\bibfnamefont {D.}~\bibnamefont
			{Riste}}, \bibinfo {author} {\bibfnamefont {M.}~\bibnamefont {Dukalski}},
		\bibinfo {author} {\bibfnamefont {C.}~\bibnamefont {Watson}}, \bibinfo
		{author} {\bibfnamefont {G.}~\bibnamefont {De~Lange}}, \bibinfo {author}
		{\bibfnamefont {M.}~\bibnamefont {Tiggelman}}, \bibinfo {author}
		{\bibfnamefont {Y.~M.}\ \bibnamefont {Blanter}}, \bibinfo {author}
		{\bibfnamefont {K.~W.}\ \bibnamefont {Lehnert}}, \bibinfo {author}
		{\bibfnamefont {R.}~\bibnamefont {Schouten}}, \ and\ \bibinfo {author}
		{\bibfnamefont {L.}~\bibnamefont {DiCarlo}},\ }\href {\doibase
		10.1038/nature12513} {\bibfield  {journal} {\bibinfo  {journal} {Nature}\
		}\textbf {\bibinfo {volume} {502}} (\bibinfo {year} {2013}),\
		10.1038/nature12513}\BibitemShut {NoStop}%
	\bibitem [{\citenamefont {Filipp}\ \emph {et~al.}(2009)\citenamefont {Filipp},
		\citenamefont {Maurer}, \citenamefont {Leek}, \citenamefont {Baur},
		\citenamefont {Bianchetti}, \citenamefont {Fink}, \citenamefont {G\"oppl},
		\citenamefont {Steffen}, \citenamefont {Gambetta}, \citenamefont {Blais},\
		and\ \citenamefont {Wallraff}}]{joint-readout}%
	\BibitemOpen
	\bibfield  {author} {\bibinfo {author} {\bibfnamefont {S.}~\bibnamefont
			{Filipp}}, \bibinfo {author} {\bibfnamefont {P.}~\bibnamefont {Maurer}},
		\bibinfo {author} {\bibfnamefont {P.~J.}\ \bibnamefont {Leek}}, \bibinfo
		{author} {\bibfnamefont {M.}~\bibnamefont {Baur}}, \bibinfo {author}
		{\bibfnamefont {R.}~\bibnamefont {Bianchetti}}, \bibinfo {author}
		{\bibfnamefont {J.~M.}\ \bibnamefont {Fink}}, \bibinfo {author}
		{\bibfnamefont {M.}~\bibnamefont {G\"oppl}}, \bibinfo {author} {\bibfnamefont
			{L.}~\bibnamefont {Steffen}}, \bibinfo {author} {\bibfnamefont {J.~M.}\
			\bibnamefont {Gambetta}}, \bibinfo {author} {\bibfnamefont {A.}~\bibnamefont
			{Blais}}, \ and\ \bibinfo {author} {\bibfnamefont {A.}~\bibnamefont
			{Wallraff}},\ }\href {\doibase 10.1103/PhysRevLett.102.200402} {\bibfield
		{journal} {\bibinfo  {journal} {Phys. Rev. Lett.}\ }\textbf {\bibinfo
			{volume} {102}},\ \bibinfo {pages} {200402} (\bibinfo {year}
		{2009})}\BibitemShut {NoStop}%
	\bibitem [{\citenamefont {Majer}\ \emph {et~al.}(2015)\citenamefont {Majer},
		\citenamefont {Chow}, \citenamefont {Gambetta}, \citenamefont {Koch},
		\citenamefont {Johnson}, \citenamefont {Schreier}, \citenamefont {Frunzio},
		\citenamefont {Schuster}, \citenamefont {Houck}, \citenamefont {Wallraff},
		\citenamefont {Blais}, \citenamefont {Devoret}, \citenamefont {Girvin},\ and\
		\citenamefont {Schoelkopf}}]{joint-readout-2D}%
	\BibitemOpen
	\bibfield  {author} {\bibinfo {author} {\bibfnamefont {J.}~\bibnamefont
			{Majer}}, \bibinfo {author} {\bibfnamefont {J.~M.}\ \bibnamefont {Chow}},
		\bibinfo {author} {\bibfnamefont {J.~M.}\ \bibnamefont {Gambetta}}, \bibinfo
		{author} {\bibfnamefont {J.}~\bibnamefont {Koch}}, \bibinfo {author}
		{\bibfnamefont {B.~R.}\ \bibnamefont {Johnson}}, \bibinfo {author}
		{\bibfnamefont {J.~A.}\ \bibnamefont {Schreier}}, \bibinfo {author}
		{\bibfnamefont {L.}~\bibnamefont {Frunzio}}, \bibinfo {author} {\bibfnamefont
			{D.~I.}\ \bibnamefont {Schuster}}, \bibinfo {author} {\bibfnamefont {A.~A.}\
			\bibnamefont {Houck}}, \bibinfo {author} {\bibfnamefont {A.}~\bibnamefont
			{Wallraff}}, \bibinfo {author} {\bibfnamefont {A.}~\bibnamefont {Blais}},
		\bibinfo {author} {\bibfnamefont {M.~H.}\ \bibnamefont {Devoret}}, \bibinfo
		{author} {\bibfnamefont {S.~M.}\ \bibnamefont {Girvin}}, \ and\ \bibinfo
		{author} {\bibfnamefont {R.~J.}\ \bibnamefont {Schoelkopf}},\ }\href
	{\doibase 10.1038/nature06184} {\bibfield  {journal} {\bibinfo  {journal}
			{Nature}\ }\textbf {\bibinfo {volume} {449}},\ \bibinfo {pages} {443}
		(\bibinfo {year} {2015})}\BibitemShut {NoStop}%
	\bibitem [{\citenamefont {Hatridge}\ \emph {et~al.}(2011)\citenamefont
		{Hatridge}, \citenamefont {Vijay}, \citenamefont {Slichter}, \citenamefont
		{Clarke},\ and\ \citenamefont {Siddiqi}}]{JPA-Hatridge}%
	\BibitemOpen
	\bibfield  {author} {\bibinfo {author} {\bibfnamefont {M.}~\bibnamefont
			{Hatridge}}, \bibinfo {author} {\bibfnamefont {R.}~\bibnamefont {Vijay}},
		\bibinfo {author} {\bibfnamefont {D.~H.}\ \bibnamefont {Slichter}}, \bibinfo
		{author} {\bibfnamefont {J.}~\bibnamefont {Clarke}}, \ and\ \bibinfo {author}
		{\bibfnamefont {I.}~\bibnamefont {Siddiqi}},\ }\href {\doibase
		10.1103/PhysRevB.83.134501} {\bibfield  {journal} {\bibinfo  {journal} {Phys.
				Rev. B}\ }\textbf {\bibinfo {volume} {83}},\ \bibinfo {pages} {134501}
		(\bibinfo {year} {2011})}\BibitemShut {NoStop}%
	\bibitem [{\citenamefont {Johansson}\ \emph {et~al.}(2012)\citenamefont
		{Johansson}, \citenamefont {Nation},\ and\ \citenamefont {Nori}}]{qutip1}%
	\BibitemOpen
	\bibfield  {author} {\bibinfo {author} {\bibfnamefont {J.}~\bibnamefont
			{Johansson}}, \bibinfo {author} {\bibfnamefont {P.}~\bibnamefont {Nation}}, \
		and\ \bibinfo {author} {\bibfnamefont {F.}~\bibnamefont {Nori}},\ }\href
	{\doibase https://doi.org/10.1016/j.cpc.2012.02.021} {\bibfield  {journal}
		{\bibinfo  {journal} {Computer Physics Communications}\ }\textbf {\bibinfo
			{volume} {183}},\ \bibinfo {pages} {1760 } (\bibinfo {year}
		{2012})}\BibitemShut {NoStop}%
	\bibitem [{\citenamefont {Johansson}\ \emph {et~al.}(2013)\citenamefont
		{Johansson}, \citenamefont {Nation},\ and\ \citenamefont {Nori}}]{qutip2}%
	\BibitemOpen
	\bibfield  {author} {\bibinfo {author} {\bibfnamefont {J.}~\bibnamefont
			{Johansson}}, \bibinfo {author} {\bibfnamefont {P.}~\bibnamefont {Nation}}, \
		and\ \bibinfo {author} {\bibfnamefont {F.}~\bibnamefont {Nori}},\ }\href
	{\doibase https://doi.org/10.1016/j.cpc.2012.11.019} {\bibfield  {journal}
		{\bibinfo  {journal} {Computer Physics Communications}\ }\textbf {\bibinfo
			{volume} {184}},\ \bibinfo {pages} {1234 } (\bibinfo {year}
		{2013})}\BibitemShut {NoStop}%
	\bibitem [{\citenamefont {Banaszek}\ \emph {et~al.}(1999)\citenamefont
		{Banaszek}, \citenamefont {D'Ariano}, \citenamefont {Paris},\ and\
		\citenamefont {Sacchi}}]{MLE}%
	\BibitemOpen
	\bibfield  {author} {\bibinfo {author} {\bibfnamefont {K.}~\bibnamefont
			{Banaszek}}, \bibinfo {author} {\bibfnamefont {G.~M.}\ \bibnamefont
			{D'Ariano}}, \bibinfo {author} {\bibfnamefont {M.~G.~A.}\ \bibnamefont
			{Paris}}, \ and\ \bibinfo {author} {\bibfnamefont {M.~F.}\ \bibnamefont
			{Sacchi}},\ }\href {\doibase 10.1103/PhysRevA.61.010304} {\bibfield
		{journal} {\bibinfo  {journal} {Phys. Rev. A}\ }\textbf {\bibinfo {volume}
			{61}},\ \bibinfo {pages} {010304} (\bibinfo {year} {1999})}\BibitemShut
	{NoStop}%
	\bibitem [{\citenamefont {James}\ \emph {et~al.}(2001)\citenamefont {James},
		\citenamefont {Kwiat}, \citenamefont {Munro},\ and\ \citenamefont
		{White}}]{MLE1}%
	\BibitemOpen
	\bibfield  {author} {\bibinfo {author} {\bibfnamefont {D.~F.~V.}\
			\bibnamefont {James}}, \bibinfo {author} {\bibfnamefont {P.~G.}\ \bibnamefont
			{Kwiat}}, \bibinfo {author} {\bibfnamefont {W.~J.}\ \bibnamefont {Munro}}, \
		and\ \bibinfo {author} {\bibfnamefont {A.~G.}\ \bibnamefont {White}},\ }\href
	{\doibase 10.1103/PhysRevA.64.052312} {\bibfield  {journal} {\bibinfo
			{journal} {Phys. Rev. A}\ }\textbf {\bibinfo {volume} {64}},\ \bibinfo
		{pages} {052312} (\bibinfo {year} {2001})}\BibitemShut {NoStop}%
	\bibitem [{\citenamefont {Chow}\ \emph {et~al.}(2009)\citenamefont {Chow},
		\citenamefont {Gambetta}, \citenamefont {Tornberg}, \citenamefont {Koch},
		\citenamefont {Bishop}, \citenamefont {Houck}, \citenamefont {Johnson},
		\citenamefont {Frunzio}, \citenamefont {Girvin},\ and\ \citenamefont
		{Schoelkopf}}]{Chow-RB-PRL}%
	\BibitemOpen
	\bibfield  {author} {\bibinfo {author} {\bibfnamefont {J.~M.}\ \bibnamefont
			{Chow}}, \bibinfo {author} {\bibfnamefont {J.~M.}\ \bibnamefont {Gambetta}},
		\bibinfo {author} {\bibfnamefont {L.}~\bibnamefont {Tornberg}}, \bibinfo
		{author} {\bibfnamefont {J.}~\bibnamefont {Koch}}, \bibinfo {author}
		{\bibfnamefont {L.~S.}\ \bibnamefont {Bishop}}, \bibinfo {author}
		{\bibfnamefont {A.~A.}\ \bibnamefont {Houck}}, \bibinfo {author}
		{\bibfnamefont {B.~R.}\ \bibnamefont {Johnson}}, \bibinfo {author}
		{\bibfnamefont {L.}~\bibnamefont {Frunzio}}, \bibinfo {author} {\bibfnamefont
			{S.~M.}\ \bibnamefont {Girvin}}, \ and\ \bibinfo {author} {\bibfnamefont
			{R.~J.}\ \bibnamefont {Schoelkopf}},\ }\href {\doibase
		10.1103/PhysRevLett.102.090502} {\bibfield  {journal} {\bibinfo  {journal}
			{Phys. Rev. Lett.}\ }\textbf {\bibinfo {volume} {102}},\ \bibinfo {pages}
		{090502} (\bibinfo {year} {2009})}\BibitemShut {NoStop}%
	\bibitem [{\citenamefont {Motzoi}\ \emph {et~al.}(2009)\citenamefont {Motzoi},
		\citenamefont {Gambetta}, \citenamefont {Rebentrost},\ and\ \citenamefont
		{Wilhelm}}]{drag-pulse}%
	\BibitemOpen
	\bibfield  {author} {\bibinfo {author} {\bibfnamefont {F.}~\bibnamefont
			{Motzoi}}, \bibinfo {author} {\bibfnamefont {J.~M.}\ \bibnamefont
			{Gambetta}}, \bibinfo {author} {\bibfnamefont {P.}~\bibnamefont
			{Rebentrost}}, \ and\ \bibinfo {author} {\bibfnamefont {F.~K.}\ \bibnamefont
			{Wilhelm}},\ }\href {\doibase 10.1103/PhysRevLett.103.110501} {\bibfield
		{journal} {\bibinfo  {journal} {Phys. Rev. Lett.}\ }\textbf {\bibinfo
			{volume} {103}},\ \bibinfo {pages} {110501} (\bibinfo {year}
		{2009})}\BibitemShut {NoStop}%
	\bibitem [{\citenamefont {Chow}\ \emph {et~al.}(2011)\citenamefont {Chow},
		\citenamefont {C\'orcoles}, \citenamefont {Gambetta}, \citenamefont
		{Rigetti}, \citenamefont {Johnson}, \citenamefont {Smolin}, \citenamefont
		{Rozen}, \citenamefont {Keefe}, \citenamefont {Rothwell}, \citenamefont
		{Ketchen},\ and\ \citenamefont {Steffen}}]{CR-gate}%
	\BibitemOpen
	\bibfield  {author} {\bibinfo {author} {\bibfnamefont {J.~M.}\ \bibnamefont
			{Chow}}, \bibinfo {author} {\bibfnamefont {A.~D.}\ \bibnamefont
			{C\'orcoles}}, \bibinfo {author} {\bibfnamefont {J.~M.}\ \bibnamefont
			{Gambetta}}, \bibinfo {author} {\bibfnamefont {C.}~\bibnamefont {Rigetti}},
		\bibinfo {author} {\bibfnamefont {B.~R.}\ \bibnamefont {Johnson}}, \bibinfo
		{author} {\bibfnamefont {J.~A.}\ \bibnamefont {Smolin}}, \bibinfo {author}
		{\bibfnamefont {J.~R.}\ \bibnamefont {Rozen}}, \bibinfo {author}
		{\bibfnamefont {G.~A.}\ \bibnamefont {Keefe}}, \bibinfo {author}
		{\bibfnamefont {M.~B.}\ \bibnamefont {Rothwell}}, \bibinfo {author}
		{\bibfnamefont {M.~B.}\ \bibnamefont {Ketchen}}, \ and\ \bibinfo {author}
		{\bibfnamefont {M.}~\bibnamefont {Steffen}},\ }\href {\doibase
		10.1103/PhysRevLett.107.080502} {\bibfield  {journal} {\bibinfo  {journal}
			{Phys. Rev. Lett.}\ }\textbf {\bibinfo {volume} {107}},\ \bibinfo {pages}
		{080502} (\bibinfo {year} {2011})}\BibitemShut {NoStop}%
	\bibitem [{\citenamefont {Chen}\ \emph {et~al.}(2012)\citenamefont {Chen},
		\citenamefont {Sank}, \citenamefont {O'Malley}, \citenamefont {White},
		\citenamefont {Barends}, \citenamefont {Chiaro}, \citenamefont {Kelly},
		\citenamefont {Lucero}, \citenamefont {Mariantoni}, \citenamefont {Megrant},
		\citenamefont {Neill}, \citenamefont {Vainsencher}, \citenamefont {Wenner},
		\citenamefont {Yin}, \citenamefont {Cleland},\ and\ \citenamefont
		{Martinis}}]{multiplex-martinis}%
	\BibitemOpen
	\bibfield  {author} {\bibinfo {author} {\bibfnamefont {Y.}~\bibnamefont
			{Chen}}, \bibinfo {author} {\bibfnamefont {D.}~\bibnamefont {Sank}}, \bibinfo
		{author} {\bibfnamefont {P.}~\bibnamefont {O'Malley}}, \bibinfo {author}
		{\bibfnamefont {T.}~\bibnamefont {White}}, \bibinfo {author} {\bibfnamefont
			{R.}~\bibnamefont {Barends}}, \bibinfo {author} {\bibfnamefont
			{B.}~\bibnamefont {Chiaro}}, \bibinfo {author} {\bibfnamefont
			{J.}~\bibnamefont {Kelly}}, \bibinfo {author} {\bibfnamefont
			{E.}~\bibnamefont {Lucero}}, \bibinfo {author} {\bibfnamefont
			{M.}~\bibnamefont {Mariantoni}}, \bibinfo {author} {\bibfnamefont
			{A.}~\bibnamefont {Megrant}}, \bibinfo {author} {\bibfnamefont
			{C.}~\bibnamefont {Neill}}, \bibinfo {author} {\bibfnamefont
			{A.}~\bibnamefont {Vainsencher}}, \bibinfo {author} {\bibfnamefont
			{J.}~\bibnamefont {Wenner}}, \bibinfo {author} {\bibfnamefont
			{Y.}~\bibnamefont {Yin}}, \bibinfo {author} {\bibfnamefont {A.~N.}\
			\bibnamefont {Cleland}}, \ and\ \bibinfo {author} {\bibfnamefont {J.~M.}\
			\bibnamefont {Martinis}},\ }\href {\doibase 10.1063/1.4764940} {\bibfield
		{journal} {\bibinfo  {journal} {Applied Physics Letters}\ }\textbf {\bibinfo
			{volume} {101}},\ \bibinfo {eid} {182601} (\bibinfo {year}
		{2012})}\BibitemShut {NoStop}%
	\bibitem [{\citenamefont {Mutus}\ \emph {et~al.}(2014)\citenamefont {Mutus},
		\citenamefont {White}, \citenamefont {Barends}, \citenamefont {Chen},
		\citenamefont {Chen}, \citenamefont {Chiaro}, \citenamefont {Dunsworth},
		\citenamefont {Jeffrey}, \citenamefont {Kelly}, \citenamefont {Megrant},
		\citenamefont {Neill}, \citenamefont {O'Malley}, \citenamefont {Roushan},
		\citenamefont {Sank}, \citenamefont {Vainsencher}, \citenamefont {Wenner},
		\citenamefont {Sundqvist}, \citenamefont {Cleland},\ and\ \citenamefont
		{Martinis}}]{impa}%
	\BibitemOpen
	\bibfield  {author} {\bibinfo {author} {\bibfnamefont {J.~Y.}\ \bibnamefont
			{Mutus}}, \bibinfo {author} {\bibfnamefont {T.~C.}\ \bibnamefont {White}},
		\bibinfo {author} {\bibfnamefont {R.}~\bibnamefont {Barends}}, \bibinfo
		{author} {\bibfnamefont {Y.}~\bibnamefont {Chen}}, \bibinfo {author}
		{\bibfnamefont {Z.}~\bibnamefont {Chen}}, \bibinfo {author} {\bibfnamefont
			{B.}~\bibnamefont {Chiaro}}, \bibinfo {author} {\bibfnamefont
			{A.}~\bibnamefont {Dunsworth}}, \bibinfo {author} {\bibfnamefont
			{E.}~\bibnamefont {Jeffrey}}, \bibinfo {author} {\bibfnamefont
			{J.}~\bibnamefont {Kelly}}, \bibinfo {author} {\bibfnamefont
			{A.}~\bibnamefont {Megrant}}, \bibinfo {author} {\bibfnamefont
			{C.}~\bibnamefont {Neill}}, \bibinfo {author} {\bibfnamefont {P.~J.~J.}\
			\bibnamefont {O'Malley}}, \bibinfo {author} {\bibfnamefont {P.}~\bibnamefont
			{Roushan}}, \bibinfo {author} {\bibfnamefont {D.}~\bibnamefont {Sank}},
		\bibinfo {author} {\bibfnamefont {A.}~\bibnamefont {Vainsencher}}, \bibinfo
		{author} {\bibfnamefont {J.}~\bibnamefont {Wenner}}, \bibinfo {author}
		{\bibfnamefont {K.~M.}\ \bibnamefont {Sundqvist}}, \bibinfo {author}
		{\bibfnamefont {A.~N.}\ \bibnamefont {Cleland}}, \ and\ \bibinfo {author}
		{\bibfnamefont {J.~M.}\ \bibnamefont {Martinis}},\ }\href {\doibase
		http://dx.doi.org/10.1063/1.4886408} {\bibfield  {journal} {\bibinfo
			{journal} {Applied Physics Letters}\ }\textbf {\bibinfo {volume} {104}},\
		\bibinfo {eid} {263513} (\bibinfo {year} {2014})}\BibitemShut {NoStop}%
	\bibitem [{\citenamefont {Macklin}\ \emph {et~al.}(2015)\citenamefont
		{Macklin}, \citenamefont {O'Brien}, \citenamefont {Hover}, \citenamefont
		{Schwartz}, \citenamefont {Bolkhovsky}, \citenamefont {Zhang}, \citenamefont
		{Oliver},\ and\ \citenamefont {Siddiqi}}]{TWPA-Berkeley}%
	\BibitemOpen
	\bibfield  {author} {\bibinfo {author} {\bibfnamefont {C.}~\bibnamefont
			{Macklin}}, \bibinfo {author} {\bibfnamefont {K.}~\bibnamefont {O'Brien}},
		\bibinfo {author} {\bibfnamefont {D.}~\bibnamefont {Hover}}, \bibinfo
		{author} {\bibfnamefont {M.~E.}\ \bibnamefont {Schwartz}}, \bibinfo {author}
		{\bibfnamefont {V.}~\bibnamefont {Bolkhovsky}}, \bibinfo {author}
		{\bibfnamefont {X.}~\bibnamefont {Zhang}}, \bibinfo {author} {\bibfnamefont
			{W.~D.}\ \bibnamefont {Oliver}}, \ and\ \bibinfo {author} {\bibfnamefont
			{I.}~\bibnamefont {Siddiqi}},\ }\href {\doibase 10.1126/science.aaa8525}
	{\bibfield  {journal} {\bibinfo  {journal} {Science}\ } (\bibinfo {year}
		{2015}),\ 10.1126/science.aaa8525}\BibitemShut {NoStop}%
	\bibitem [{\citenamefont {White}\ \emph {et~al.}(2015)\citenamefont {White},
		\citenamefont {Mutus}, \citenamefont {Hoi}, \citenamefont {Barends},
		\citenamefont {Campbell}, \citenamefont {Chen}, \citenamefont {Chen},
		\citenamefont {Chiaro}, \citenamefont {Dunsworth}, \citenamefont {Jeffrey},
		\citenamefont {Kelly}, \citenamefont {Megrant}, \citenamefont {Neill},
		\citenamefont {O'Malley}, \citenamefont {Roushan}, \citenamefont {Sank},
		\citenamefont {Vainsencher}, \citenamefont {Wenner}, \citenamefont
		{Chaudhuri}, \citenamefont {Gao},\ and\ \citenamefont
		{Martinis}}]{TWPA-martinis}%
	\BibitemOpen
	\bibfield  {author} {\bibinfo {author} {\bibfnamefont {T.~C.}\ \bibnamefont
			{White}}, \bibinfo {author} {\bibfnamefont {J.~Y.}\ \bibnamefont {Mutus}},
		\bibinfo {author} {\bibfnamefont {I.-C.}\ \bibnamefont {Hoi}}, \bibinfo
		{author} {\bibfnamefont {R.}~\bibnamefont {Barends}}, \bibinfo {author}
		{\bibfnamefont {B.}~\bibnamefont {Campbell}}, \bibinfo {author}
		{\bibfnamefont {Y.}~\bibnamefont {Chen}}, \bibinfo {author} {\bibfnamefont
			{Z.}~\bibnamefont {Chen}}, \bibinfo {author} {\bibfnamefont {B.}~\bibnamefont
			{Chiaro}}, \bibinfo {author} {\bibfnamefont {A.}~\bibnamefont {Dunsworth}},
		\bibinfo {author} {\bibfnamefont {E.}~\bibnamefont {Jeffrey}}, \bibinfo
		{author} {\bibfnamefont {J.}~\bibnamefont {Kelly}}, \bibinfo {author}
		{\bibfnamefont {A.}~\bibnamefont {Megrant}}, \bibinfo {author} {\bibfnamefont
			{C.}~\bibnamefont {Neill}}, \bibinfo {author} {\bibfnamefont {P.~J.~J.}\
			\bibnamefont {O'Malley}}, \bibinfo {author} {\bibfnamefont {P.}~\bibnamefont
			{Roushan}}, \bibinfo {author} {\bibfnamefont {D.}~\bibnamefont {Sank}},
		\bibinfo {author} {\bibfnamefont {A.}~\bibnamefont {Vainsencher}}, \bibinfo
		{author} {\bibfnamefont {J.}~\bibnamefont {Wenner}}, \bibinfo {author}
		{\bibfnamefont {S.}~\bibnamefont {Chaudhuri}}, \bibinfo {author}
		{\bibfnamefont {J.}~\bibnamefont {Gao}}, \ and\ \bibinfo {author}
		{\bibfnamefont {J.~M.}\ \bibnamefont {Martinis}},\ }\href {\doibase
		http://dx.doi.org/10.1063/1.4922348} {\bibfield  {journal} {\bibinfo
			{journal} {Applied Physics Letters}\ }\textbf {\bibinfo {volume} {106}},\
		\bibinfo {eid} {242601} (\bibinfo {year} {2015})}\BibitemShut {NoStop}%
	\bibitem [{\citenamefont {Roy}\ \emph {et~al.}(2015)\citenamefont {Roy},
		\citenamefont {Kundu}, \citenamefont {Chand}, \citenamefont {Vadiraj},
		\citenamefont {Ranadive}, \citenamefont {Nehra}, \citenamefont {Patankar},
		\citenamefont {Aumentado}, \citenamefont {Clerk},\ and\ \citenamefont
		{Vijay}}]{BBparamp}%
	\BibitemOpen
	\bibfield  {author} {\bibinfo {author} {\bibfnamefont {T.}~\bibnamefont
			{Roy}}, \bibinfo {author} {\bibfnamefont {S.}~\bibnamefont {Kundu}}, \bibinfo
		{author} {\bibfnamefont {M.}~\bibnamefont {Chand}}, \bibinfo {author}
		{\bibfnamefont {A.}~\bibnamefont {Vadiraj}}, \bibinfo {author} {\bibfnamefont
			{A.}~\bibnamefont {Ranadive}}, \bibinfo {author} {\bibfnamefont
			{N.}~\bibnamefont {Nehra}}, \bibinfo {author} {\bibfnamefont {M.~P.}\
			\bibnamefont {Patankar}}, \bibinfo {author} {\bibfnamefont {J.}~\bibnamefont
			{Aumentado}}, \bibinfo {author} {\bibfnamefont {A.}~\bibnamefont {Clerk}}, \
		and\ \bibinfo {author} {\bibfnamefont {R.}~\bibnamefont {Vijay}},\ }\href
	{\doibase 10.1063/1.4939148} {\bibfield  {journal} {\bibinfo  {journal}
			{Applied Physics Letters}\ }\textbf {\bibinfo {volume} {107}},\ \bibinfo
		{pages} {262601} (\bibinfo {year} {2015})}\BibitemShut {NoStop}%
\end{thebibliography}

%

\end{document}